\documentclass[a4paper,fleqn,usenatbib,twocolumn]{mnras}

\usepackage{newtxtext,newtxmath}

\usepackage[T1]{fontenc}
\usepackage{ae,aecompl}

\usepackage{pdflscape}
\usepackage{graphicx}	
\usepackage{amsmath}	
\usepackage{amssymb}	
\usepackage{tablefootnote}

\defcitealias{Gwinn2019}{Paper\ I}

\title[Analytical Noodle Model]{Scintillation Arc Brightness and Electron Density for an Analytical Noodle Model}

\author[C. R. Gwinn and E. B. Sosenko]{
Carl R. Gwinn$^{1}$\thanks{E-mail: cgwinn@ucsb.edu (CRG)} and Evan B. Sosenko$^{1,2}$\\
$^{1}$Physics Department, University of California, Santa Barbara, California 93106, USA\\
$^{2}$ Physics Department, University of California, Riverside, California 92521, USA
}

\date{Accepted XXX. Received YYY; in original form ZZZ}

\pubyear{2018}

\begin{document}
\label{firstpage}
\pagerange{\pageref{firstpage}--\pageref{lastpage}}
\maketitle

\begin{abstract}
We show that narrow filaments or sheets of over- or under-dense plasma, or ``noodles,'' with fluctuations of scattering phase of less than a radian, can form the scintillation arcs seen for many pulsars. The required local fluctuations of electron density are indefinitely small. We assume a cosine profile for the electron column and find the scattered field by analytic Kirchhoff integration. For a large electron column, corresponding to large amplitude of phase variation, the stationary-phase approximation is accurate; we call this regime ``ray optics''. For smaller-amplitude phase variation, the stationary-phase approximation is inaccurate or inapplicable; we call this regime ``wave optics''. 
We show that scattering is most efficient when the width of the strip equals that of one pair of Fresnel zones, and in the wave-optics regime. We show that the resolution of present observations is about 100 Fresnel zones on the scattering screen.  Incoherent superposition of strips within a resolution element tends to increase the scattered field. We find that observations match a single noodle per resolution element with phase of up to 12 radians; or many noodles per resolution element with arbitrarily small phase variation each, for net phase of less than a radian. Observations suggest a minimum radius for noodles of about 650\ km, comparable to the ion inertial scale or the ion cyclotron radius in the scattering plasma.
\end{abstract}




\section{Introduction}

Scattering of nearby pulsars often appears as scintillation arcs. These structures take their name from their arc-like appearance in secondary spectrum. The secondary spectrum $C(\tau,f)$ is the square modulus of intensity $I(\tau,f)$, in the domain of delay $\tau$ and rate $f$. This domain is the 2D Fourier conjugate of the domain of frequency $\nu$ and time $t$. The intensity $\tilde I(\nu,t)$ in the frequency-time domain is known as the dynamic spectrum. A point in the secondary spectrum appears in the dynamic spectrum $\tilde I(\nu,t)$ as a sinusoidal variation of intensity with frequency $\nu$, with a linear shift in time $t$ of the phase of the sinusoid.  For a pulsar that shows scintillation arcs, the superposition of many such sinusoids leads to many points, organized along a pair of parabolic arcs in the secondary spectrum. Apexes of the parabolas lie at the origin, and axes along the positive and negative delay coordinate axis \citep{2001ApJ...549L..97S,2003ApJ...599..457H,2006ChJAS...6b.233P,2007A&AT...26..517S,2007ASPC..365..254S,2010ApJ...708..232B}. In some high-sensitivity observations, secondary ``arclets'' appear, with apexes along the primary arc, and with the same curvature as the primary arc but opening in the opposite direction. 
The extent of scintillation arcs, to high delay and rate, makes clear that they are a result of interstellar scattering, with deflections of up to 10\ AU by material at distances of about 100 parsecs. The net intensity of the arc is no more than a few percent of the total, indicating that such scattering is relatively rare. 

In a previous paper \citep[][hereafter ``Paper\ I'']{Gwinn2019}, we considered a model for scintillation arcs consisting of long, straight, narrow strips of refracting material: the ``noodle'' model. 
Physically, the noodles may take the form of filaments (``spaghetti'') or sheets (``lasagna'').
Magnetized plasmas often contain over- or under-dense sheets or filaments \citep{1989JOSAA...6..977C,1990ApJ...358..685A,1994Natur.372..754D,1997JGR...102..263G,2019ApJ...878..157X}
These density variations give rise to changes in refractive index. 
Such structures are expected to appear at small scales in reconnection sheets, 
as discussed in detail in Section\ 1.3 of \citetalias{Gwinn2019}.

We suppose that the scattering region is relatively concentrated along the line of sight, so that the change in phase from refraction by the noodle can be projected onto a strip in a thin screen. A Kirchhoff integral yields the scattered field. The phase varies across the strip with some profile $\varphi(x)$. The noodle model assumes that profile remains the same along the length of the strip. \citetalias{Gwinn2019} showed that an assemblage of such strips, parallel and at different lateral distances from the undeflected line of sight, gives rise to scintillation arcs. 

\begin{figure*}
\centering
\includegraphics[width=0.98\textwidth]{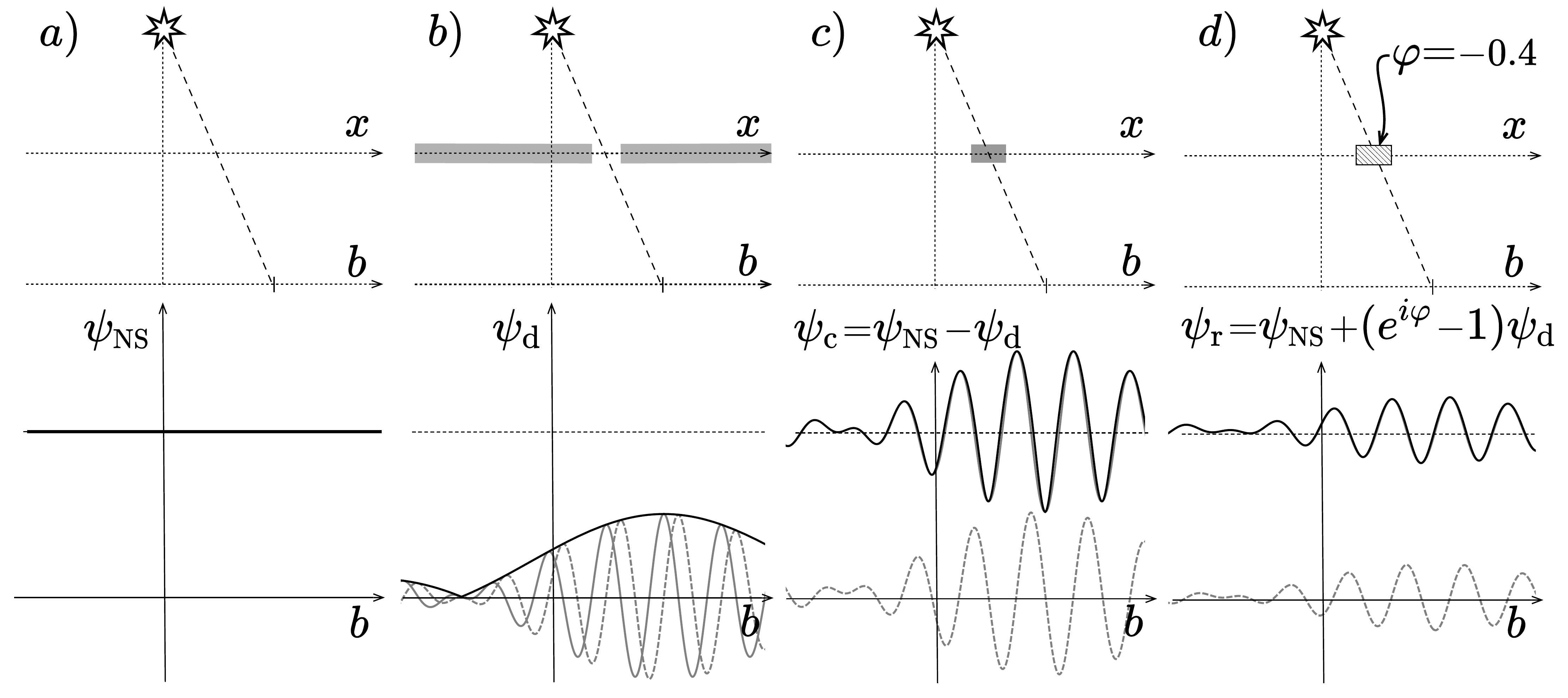} 
\caption{Thought experiment showing wave optics of the noodle model. 
A point source illuminates a screen, and an observer measures the diffracted field as a function of position.
If the screen is absent (panel $a$ at left), the observer measures uniform field $\psi_\mathrm{NS}$.
If the screen is opaque except for a slit displaced from the optical axis ($b$), the observer measures a diffracted field $\psi_\mathrm{d}$ with phase oscillation corresponding to that displacement: gray lines show real (solid) and imaginary (dashed) components, and black line shows magnitude.
If the screen is transparent except for an opaque strip at the position of the slit ($c$), the field $\psi_\mathrm{c}$ is the difference of the fields for $a$ and $b$.
If instead a thin strip of transparent refracting material replaces the slit ($d$), the field $\psi_\mathrm{r}$ is the sum of $\psi_\mathrm{d}$ and a phase-shifted copy of $\psi_\mathrm{c}$.
Real part and amplitude are nearly indistinguishable in the latter two cases.
\label{fig:ThoughtExperiment}}
\end{figure*}

\subsection{Outline of Paper}

In this paper, we consider the properties of scintillation arcs for a simple model profile for the phase variation across a strip $\varphi(x)$, and compare results with observations. Our model allows us to calculate and compare brightnesses of features in the secondary spectrum, for different strip widths $w$ and phase amplitude $\varphi_0$, at different offsets from the line of sight $x_j$. In Section\ \ref{sec:OpticalFramework} we briefly recapitulate the optical framework for the calculation, using Kirchhoff diffraction, as presented in more detail in \citetalias{Gwinn2019}.
We calculate the brightness of the resulting feature on the primary arc, and show that it is proportional to the squared magnitude of the Kirchhoff integral.

We present our model profile $\varphi(x)$ in Section\ \ref{sec:CosineStripModel}. 
In our model, $\varphi(x)$ follows one wavelength of a cosine, offset so that the phase and its first derivative fall to zero at either edge.
The parameters of the model are the amplitude of the cosine $\varphi_0$, proportional to the maximum electron column of the strip; the cosine period $w$, equal to the width of the strip;
and the offset of the strip at its closest approach to the optical axis $x_j$.
If the width of the strip is an integral number of Fresnel zones, the Kirchhoff integral across the strip yields a Bessel function. If not, it can be computed numerically. 
Our calculations show that the strip scatters most efficiently when its width is equal to one pair of Fresnel zones, as we discuss in Section\ \ref{sec:NumericalExtensionAndDiscussion}.
We find the Kirchhoff integral in the limit of screen phase less than one radian, $\varphi_0\ll 1$, in Section\ \ref{sec:SmallElectronColumn}.
In Section\ \ref{sec:StatPhaseApprox}, we show that in the large-amplitude limit, the stationary phase approximation gives the integral across the strip, and the ray approximation describes the scattering. We call this the ``ray'' limit. In the limit of smaller phase amplitude, the stationary-phase approximation fails. We call this the ``wave'' limit. 
The model thus provides a good illustration of the stationary-phase approximation and its limitations.

We discuss the time evolution of the field from a model strip, as the undeflected line of sight approaches it, passes through, and moves away, in Section\ \ref{sec:TimeEvolution}.
A model strip appears only within a certain offset of the undeflected line of sight; this maximum offset is one pair of Fresnel zones for a strip with $|\varphi_0| \le 1$,
and $n\lesssim |\varphi_0|$ pairs of Fresnel zones for $|\varphi_0|>1$.
This provides a natural selection effect that concentrates scattered images near the undeflected line of sight. For linear, parallel strips, it maintains an elongated distribution of scattered images centered on the pulsar, even as the pulsar moves across the sky, as observed.
We compare different model strips at the same offset in Section\ \ref{sec:BrightnessAtOneXj}.
In Section\ \ref{sec:ObservationalResolution} we apply results of\ \citetalias{Gwinn2019} to show that present observations have a spatial resolution of about 100 pairs of Fresnel zones at the screen; they cannot distinguish between strips closer than this separation. 
As we discuss in Section\ \ref{sec:SuperpositionOfStrips}, multiple strips within a resolution element combine incoherently, to produce brighter points and arclets in the secondary spectrum. 

We compare results of our calculations with the observed brightnesses of scintillation arcs in Section\ \ref{sec:Observations}, and find our models easily reproduce the observations. We compute the screen phase required to reproduce the observed intensities, and find that $\varphi_0$ is a few radians for a single strip within an observational resolution element, but can be much less -- indeed, arbitrarily small -- for multiple incoherently-superposed strips within one resolution element. We calibrate $\varphi_0$ in electron column in Section\ \ref{sec:ElectronDenstiyVariationsFromNoodles}. 
We find that the net variations in electron column, in the screen, can be as small as $10^{10}\ \mathrm{cm}^{-2}$; the electron column of one noodle can be arbitrarily small.
In Section\ \ref{sec:InnerScale}, we suggest that the maximum observed delay along the arc corresponds to a minimum noodle width of about 650\ km. This is about the ion cyclotron radius, or the ion inertial scale, in the scattering material; similar dimensions have been proposed for an inner scale of scattering \citep{1990ApJ...353L..29S,2018ApJ...865..104J,2019ApJ...878..157X}.
In Section\ \ref{sec:ComparisonWithOtherModels}, we briefly compare our results with those of ``sheet'' and ``spheroidal'' models. Previous sheet models are closely related to the noodle model; they differ in that they use ray optics, and do not appeal to incoherent superposition. In spheroidal models, scattering arises from a single, fixed point of stationary phase on the screen. Such models require much greater electron column than noodles to attain the same brightness in the secondary spectrum. Spheroidal models also require some external mechanism to orient the scatterers relative to the line of sight and to maintain an elongated distribution along a constant direction on the sky.

\subsection{Thought Experiment}

A major puzzle in understanding scintillation arcs is the enormous variation of electron column densities over tiny lateral length scales that their scattering appears to require.
These large column densities arise from the large delays seen for scattering, implying a large angular deflection, and consequently, in the context of ray optics, a large phase gradient at the scatterer.
To sustain that gradient across 
the size of the scatterer, as inferred from its brightness, requires a large electron column on one side of the scatterer relative to the other. 
For roughly spherical scatterers as deep along the line of sight as they are wide across it, the local fluctuation of electron density easily exceeds $10^2\ \mathrm{cm}^{-3}$,
far greater than the typical electron density of $0.25\ \mathrm{cm}^{-3}$ in the scattering plasma \citep{1990ApJ...353L..29S}.

In this paper, we show that wave optics can produce scintillation arcs with much lower electron column than ray optics.
A thought experiment, depicted in Figure\ \ref{fig:ThoughtExperiment}, illustrates the principle. 
A sufficiently narrow slit in an opaque screen deflects light through large angles, leading to a diffracted field in the observer plane $\psi_\mathrm{d}$.
If the slit is offset from the optical axis, this field shows large phase variations;
a second slit, as in Young's two-slit experiment, will show these phase variations via interference.
On the other hand, an opaque strip at the slit position, and no screen elsewhere, will lead to the complementary field $\psi_\mathrm{c}$:
this is the field with no screen anywhere, minus the response of the strip: $\psi_\mathrm{c} = \psi_\mathrm{NS} - \psi_\mathrm{d}$.
However, 
if the slit is replaced by a narrow strip of refracting material, and the screen is removed, then diffraction from the strip will interfere with the undeflected field to produce variations in intensity at the observer. 
Mathematically, this will be the sum of three parts: the response with no screen, minus the response of the slit, plus the response of the slit times the phase introduced by the refracting strip:
$\psi_\mathrm{r} = \psi_\mathrm{NS} + \left( e^{i \varphi} -1 \right) \psi_\mathrm{d}$.
Phase variations from the offset of the refracting strip lead to an interference pattern at the observer.
Sections\ \ref{sec:OpticalFramework} and\ \ref{sec:CosineStripModel} present this argument formally, using Kirchhoff integration.

Incoherent superposition of nearby strips serves to further reduce the electron column required to produce the observed brightness of the secondary spectrum.
The observational resolution of the secondary spectrum, expressed as a linear resolution at the scattering screen, is limited. For present observations, strips within about 100 pairs of Fresnel zones combine to produce a single observed feature in the secondary spectrum.
Using wave optics and superposition,
we find that many such strips, with an arbitrarily small electron column, can reproduce observed brightnesses of scintillation arcs.

Observational studies suggest, and interferometric observations confirm, that scatterers are concentrated near the source on the sky, in an elongated distribution \citep{2004MNRAS.354...43W,2010ApJ...708..232B}. 
The anisotropy requires aligned anisotropic scatterers, with communication or some external agency to reinforce their alignment.
As noted in \citetalias{Gwinn2019}, a large-scale magnetic field is a good candidate for an aligning agent.
Plasma fluctuations tied to lines of magnetic field provide a natural way to form elongated noodles,
as discussed in Section\ 1.3 of \citetalias{Gwinn2019}.
A large-scale magnetic field with the lowest-energy configuration has constant direction and magnitude, and the resulting field-aligned density fluctuations will be parallel.
For the noodle model, the segment of the noodle that is closest to the undeflected line of sight will shift with motions of source and observer, so as to move with the undeflected line of sight.
This segment plays the role of the scatterer that the observer tracks over time, as discussed in Section\ 2 of \citetalias{Gwinn2019} and in Section\ \ref{sec:ComparisonWithOtherModels} below.
If the noodles are parallel, or nearly parallel, this preserves the anisotropy of the distribution, while ensuring that the scatterers remain near the source.

\section{Optical Framework}\label{sec:OpticalFramework}

We briefly describe the fundamentals of our model in this section;  \citetalias{Gwinn2019}
 gives details. We argue that the scattering may be modeled as a thin screen, somewhere along the line of sight \citepalias[see][Section\ 5.1]{Gwinn2019}. We suppose that the source is pointlike, and that source and observer move at constant velocity relative to a static pattern of phase variations in the screen, imposed by variations electron column. We use Kirchhoff diffraction theory, and calculate the field as a function of frequency, time, and position in the observer plane.

\subsection{Screen Phase from Plasma Propagation}\label{sec:PlasmaPhase}

Propagation through a medium with a refractive index introduces a phase additional to that from propagation through vacuum.
The refractive index of a plasma is related to the plasma frequency $\omega_\mathrm{p}$:
\begin{align}
\omega_\mathrm{p}=\sqrt{4\pi r_0 c^2 N_e}
,
\end{align}
where $N_e$ is the number density of electrons and $r_0$ is the classical electron radius $r_0 ={e^2}/{m c^2}$,
$e$ and $m$ are the electron charge and mass,
and $c$ is the speed of light.
For observing frequency $\nu = \omega /2\pi$, in the limit $\omega \gg \omega_\mathrm{p}$ where $\omega_p$ is the plasma frequency,
a column of plasma introduces the phase:
 \begin{align}
\label{eq:basic_plasma_phase}
\varphi ( \nu )&\approx - \frac{c r_0}{\nu} \int_0^L  dz\, N_e
.
\end{align}
The integrated electron density along the path $\int N_\mathrm{e}dz$ is known as electron column.
The negative sign indicates that the phase introduced by the plasma is negative, because the index of refraction is less than 1.
In this paper, we are concerned with the incremental electron column introduced by one scattering structure, $\int_0^L dz\, \Delta N_e$,
where $\Delta N_e$ is the fluctuation in electron density of one path through the screen, relative to the average over the screen.
As a specific example, used as a fiducial value later in this paper, a phase change of $\varphi=2\ \mathrm{radians}$ at an observing frequency of $\nu=322.495\ \mathrm{MHz}$ corresponds to an electron column of $\int_0^L dz\, \Delta N_\mathrm{e}= - 7.9\times 10^{10}\ \mathrm{cm}^{-2}$.

\subsection{Kirchhoff Integral}\label{sec:KirchhoffIntegral}

\begin{figure}
\centering
\includegraphics[width=0.45\textwidth]{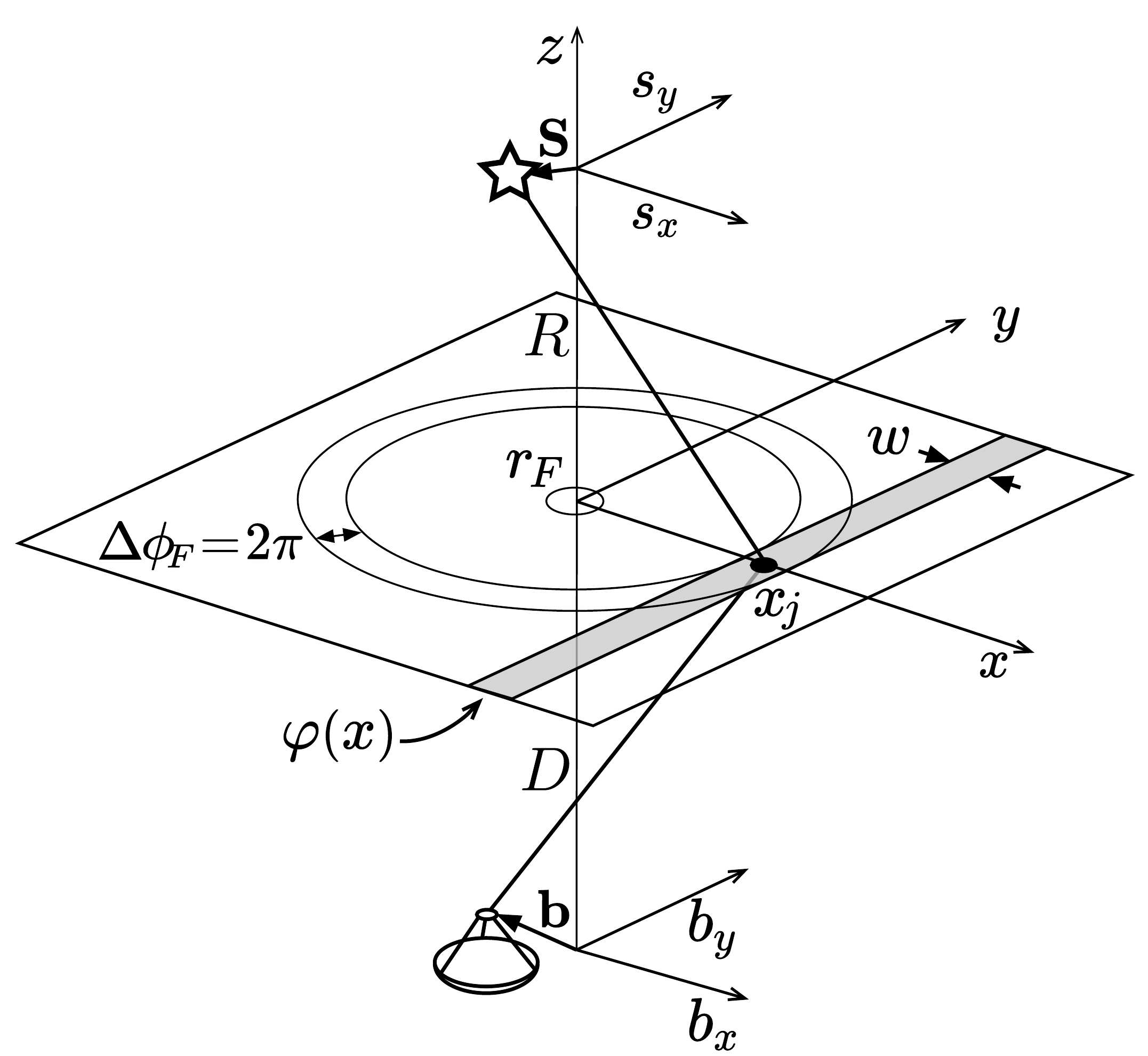} 
\caption{Geometry for Kirchhoff integration.
The optical axis is vertical.
The observer plane is $\mathbf{b}$ and the source plane is $\mathbf{s}$.
Scattering material lies in a thin screen $\mathbf{x}$, in this case a narrow strip of width $w$ at offset $x_j$ from the optical axis. 
The Fresnel phase $\phi_\mathrm{F}$ is constant on circles around the optical axis in the screen plane.
\label{fig:ReferenceGeom}}
\end{figure}

For a point source at that emits field $\psi_\mathrm{src}$, the Kirchhoff integral takes the form \citepalias{Gwinn2019}:
\begin{align}
\label{eq:KirchhoffIntegral}
\psi_\mathrm{obs} (\mathbf{b}) &= \frac{-i}{2\pi r_\mathrm{F}^2} \int d^2x\; e^{i\left[ \left( \frac{k}{2 D}\right) |\mathbf{b} - \mathbf{x} |^2 + \varphi(\mathbf{x}) \right]} 
 e^{i\left( \frac{k}{2R}\right) |\mathbf{x} - \mathbf{s} |^2} \psi_\mathrm{src} 
,
\end{align}
where the lateral coordinates are $\mathbf{s}$ in the source plane, $\mathbf{x}$ in the screen plane, and $\mathbf{b}$ in the observer plane; in each of these planes, the coordinate origin lies on the optical axis.
Relative to that axis, the source lies at $\mathbf{s} = (s_x,s_y)$ and the observer at $\mathbf{b} = (b_x,b_y)$.
Figure\ \ref{fig:ReferenceGeom} shows the geometry.
The integral is over the screen, and $\varphi(\mathbf{x})$ is the phase introduced at the screen by plasma refraction.
The distance from source to screen is $R$, and from screen to observer is $D$.
The magnification of the scattering screen treated as a lens is:
\begin{align}
M &= \frac{D}{R} 
\end{align}
Note that the Kirchhoff integral in\ \ref{eq:KirchhoffIntegral}, without the normalization, has units of area.
The contribution to the field of the Kirchhoff integral, or of the integral over part of the screen, is proportional to that effective area.
The normalization factor $-i/2\pi r_\mathrm{F}^2$ ensures that 
the observed field $\psi_\mathrm{obs}$ and the source field $\psi_\mathrm{src}$ have the same dimensions.

The Fresnel scale is the fundamental scale of the problem:
\begin{align}
r_\mathrm{F} &=\sqrt{ \frac{DR}{(D+R)} \frac{1}{k} } 
= \sqrt{\frac{1}{(1+M)}\frac{D}{k}} 
.
\end{align}
The Fresnel phase is:
\begin{align}
\phi_\mathrm{F} &=\frac{|\mathbf{x}|^2}{2 r_\mathrm{F}^2}
\end{align}
The Fresnel phase corresponds to the geometric length of a path that passes from a source on the optical axis to an observer on the axis, and through a point on the screen at $\mathbf{x}$ from that axis.  
Observations of the scintillation arcs indicate that the scatterers lie at least as far as $x \approx 1.4\times 10^3\, r_\mathrm{F}$ from the undeflected line of sight, as we discuss in Section\ \ref{sec:FormationFigure}.
A ``Fresnel zone'' is the annulus between $\phi_\mathrm{F}=(N+\frac{1}{2})\pi$ and $\phi_\mathrm{F}=(N+\frac{3}{2})\pi$, where $N$ is an integer (except for the first Fresnel zone, which extends from $\phi_\mathrm{F}=0$ to $\phi_\mathrm{F}=\frac{1}{2}\pi$). Thus, across a pair of adjacent Fresnel zones, the Fresnel phase changes by $2\pi$:
\begin{align}
\label{eq:W2PiDef}
W_{2\pi}
\equiv  2\pi \left( \frac {\partial \phi_\mathrm{F}}{\partial x}\right)^{-1} = 2\pi \frac{r_\mathrm{F}^2}{x}
\end{align}
\label{eq:WidthOf1PairOfFresnelZones}
where $W_{2\pi}$ is the width on the screen of one pair of adjacent Fresnel zones.
The width of $n$ pairs of Fresnel zones at a distance $x$ from the origin is approximately:
\begin{align}
\label{eq:WidthOfnPairsOfFresnelZones}
w =
n \cdot W_{2\pi}
.
\end{align}

\subsection{Narrow Strip}\label{sec:NarrowStrip}

We consider the scattering effect of a narrow strip of phase-changing material, where the screen phase $\varphi(x)$ varies across the strip, with a profile that is unchanging along its length. Figure\ \ref{fig:ReferenceGeom} shows such a strip.
We suppose that the strip is displaced from the undeflected line of sight along the $x$ axis, with the midline of the strip at $x_j$. The width of the strip is $w$.

We make the realistic assumptions discussed in \citepalias{Gwinn2019}: 
the strip is many times the Fresnel scale from the undeflected line of sight, the strip is narrower than a Fresnel scale, and source and observer do not move more than about a Fresnel scale during an observation. We also suppose that the observed frequency range is small compared with the frequency at the center of that range.

The net observed field at the observer is:
\begin{align}
\label{eq:DiffractiveStripSimplified}
\psi_\mathrm{obs} 
&\approx \psi_\mathrm{NS}   + \psi_{\mathrm{s}j} -  \psi_{\mathrm{s}0j}
\end{align}
where $\psi_\mathrm{NS}$ is the field in the absence of any scattering material in the screen -- that is, with zero screen phase;
$\psi_{\mathrm{s}j}$ is the field from the strip centered at $x_j$, and $\psi_{\mathrm{s}j0}$ is the contribution of that strip with zero screen phase.
As discussed in detail in Section\ 2.2 of \citetalias{Gwinn2019}, in the absence of scattering material the un-normalized integral in\ \ref{eq:KirchhoffIntegral} yields $i 2\pi r_\mathrm{F}^2$,
with magnitude equal to the effective area $A_\mathrm{NS}=2\pi r_\mathrm{F}^2$.
This area corresponds to the undeflected line of sight.
After normalization, $\psi_\mathrm{NS} = \psi_\mathrm{src}$ .

Over the domain of the strip, the Kirchhoff integral, \ref{eq:KirchhoffIntegral}, then gives the field from the strip $\psi_{\mathrm{s}j}$ as:
\begin{align}
\label{eq:FieldFromStrip}
\psi_{\mathrm{s}j} &= \frac{-i}{{2\pi} r_\mathrm{F}^2} \left(  e^{ i \phi_{\mathrm{g}j}}  \int\limits_{-w/2}^{+w/2} du\, e^{\frac {i}{r_\mathrm{F}^2} x_j u }  e^{i\varphi_j (u)}   \right) \sqrt{2\pi i\,}\,r_\mathrm{F}\, \psi_\mathrm{src} 
,
\end{align}
where $u=x-x_j$ takes the place of the coordinate $x$.
We neglect the small curvature of the Fresnel phase across the narrow strip, as discussed in Section\ 3.2.2 of \citetalias{Gwinn2019}.
The factor of $\sqrt{2\pi i\, }\,r_\mathrm{F}$ in \ref{eq:FieldFromStrip} is the integral over the $y$-coordinate in the screen plane: 
\begin{align}
\label{eq:StationaryPhaseIntegralOverY}
\left( \int_{-\infty}^ {+\infty} dy\,e^{\frac {i}{r_\mathrm{F}^2} (y-V_y t)^2 } \right) = \sqrt{2\pi i\,}\,r_\mathrm{F}
.
\end{align}
This integral may be performed by analytic continuation of an integral with small imaginary part for the variance $r_\mathrm{F}^2$; or by the stationary-phase approximation, which is exact in this case \citep{1978amms.book.....B}.
The result for the strip with no screen phase, $\psi_{\mathrm{s}0j}$, is identical to\ \ref{eq:FieldFromStrip}, but with $\varphi_j=0$.

The geometric phase 
where the strip passes closest to the undeflected line of sight, at $\mathbf{x}=(x_j,0)$, is to a good approximation \citepalias[][Section\ 3.2.2]{Gwinn2019}:
\begin{align}
\phi_{\mathrm{g}j} 
& \approx  \frac{1}{2 r_\mathrm{F0}^2 } \left( 1 + \frac{\Delta\nu}{\nu_0}\right) \left( x_j  - V_x t   \right)^2 
\nonumber \\  
\label{eq:ApproxPhiGeomDefFreqVariation}
&\approx  \frac{1}{2 r_\mathrm{F0}^2 } \left( x_j^2  - 2 x_j  V_x t  + x_j^2 \left( \frac{\Delta\nu}{\nu_0}\right) \right) 
.
\end{align}
The subscript $j$ indicates that this phase depends on $x_j$.
The center frequency of the observed band of frequencies is $\nu_0$, 
and we use the offset $\Delta\nu = \nu-\nu_0$ to express other frequencies within the band.
The observing bandwidth is $B$: $-B/2 \le \Delta\nu \le B/2$.
Recall that we assume $B \ll \nu_0$.
The position where the undeflected line of sight from observer source pierces the screen is:
\begin{align}
\mathbf{b}_\mathrm{1}=(b_{1x},b_{1y})=(\mathbf{b}+M \mathbf{s} )/(1+M)
,
\end{align}
and the corresponding velocity of that point relative to the screen is 
\begin{align}
\mathbf{V} = (V_x,V_y)=\frac{d}{dt} \mathbf{b_1}
.
\end{align}
Note that $V_x$ appears in \ref{eq:ApproxPhiGeomDefFreqVariation}.

We combine the results of\ \ref{eq:FieldFromStrip}-\ref{eq:ApproxPhiGeomDefFreqVariation} to write the contribution of the strip, minus that of its absence, as 
\begin{align}
\psi_{\mathrm{s}j} - \psi_{\mathrm{s}0j} &= \Gamma_{j} e^{i \phi_{\mathrm{g}j}} \psi_\mathrm{src}
\end{align}
where 
\begin{align}
\label{eq:GammaJDef}
\Gamma_{j} &= \sqrt{\frac{-i}{2\pi}} \frac{1}{r_\mathrm{F}} \int_{-w/2}^{w/2}du\, e^{\frac{i}{r_\mathrm{F}^2} x_j u} \left( e^{i \varphi(u)} - 1 \right)
.
\end{align}
Note that $\Gamma_{j}$ is dimensionless; it is proportional to the characteristic area for the integral over the strip, normalized by $-i/2\pi r_\mathrm{F}^2$ as in\ \ref{eq:KirchhoffIntegral}.
Integration along the strip, in the $y$-direction, yielded a factor of $\sqrt{2\pi i} r_\mathrm{F}$ (\ref{eq:StationaryPhaseIntegralOverY}).
Integration over the $x$-direction in\ \ref{eq:GammaJDef} yields another factor of length, of order $w$.
Consequently, the effective area for the un-normalized integral is about $A_j \approx \sqrt{2\pi} r_\mathrm{F} w$.
Because the strip is narrow, $w\ll r_\mathrm{F}$, the effective area covers an extremely elongated region, with length $\sqrt{2\pi} r_\mathrm{F}$ along the strip and width $w$ across it.
The area is also much smaller than that for the undeflected line of sight, $2\pi r_\mathrm{F}^2$.

We rewrite the geometric phase in \ref{eq:ApproxPhiGeomDefFreqVariation} as:
\begin{align}
\label{eq:ApproxPhiGeomDefFreqVariationUseTauF}
\phi_{\mathrm{g}j} &\approx  \frac{x_j^2}{2 r_\mathrm{F0}^2 }+ \tau_j \Delta \nu  + f_j  t 
.
\end{align}
where we define, as in 53\ of  \citetalias{Gwinn2019}: 
\begin{alignat}{3}
\label{eq:DefineAlphaBetaGamma}
\alpha &\equiv \frac{1}{4\pi r_{\mathrm{F}0}^2 \nu_0} = \frac{ (1+M) }{2 c D}, 
\qquad\qquad\qquad& 
\tau_j &\equiv \alpha x_j^2  \\
\beta &\equiv  -\frac{V_x}{2\pi r_{\mathrm{F}0}^2} =   -\frac{(1+M) }{c D} \nu_0 V_x ,  & f_j      &\equiv \beta x_j   
.
\nonumber 
\end{alignat}
Here, $\tau_j$ and $f_j$ are real variables that depend on $j$, whereas $\alpha$ and $\beta$ are real constants independent of $j$, although they depend on the geometry and parameters of the observation.  Because $x_j$ is large compared with $r_\mathrm{F}$ and the other transverse dimensions, $x_j^2/2 r_\mathrm{F0}^2 \gg 1$, and the total geometric phase is more or less random. 

As argued in \citetalias{Gwinn2019},
the net field at the observer from an assemblage of narrow strips is:
\begin{align}
\psi_{\rm obs}&= \psi_\mathrm{NS} + \sum_j \left( \psi_{\mathrm{s}j} - \psi_{\mathrm{s}0j} \right) 
\\
&
= \psi_\mathrm{src} \left[ 1 + \sum_j \Gamma_{j} e^{i \phi_{\mathrm{g}j}} \right] 
,
\end{align}
where the sum runs over the strips. The intensity at the observer is:
\begin{align}
\label{eq:AnalyticIntensity}
I_{\rm obs}&=\psi_{\rm obs} \psi^*_{\rm obs} \\
&= \Bigg[ 1   - \sum_j 2  \Gamma_{j}  \cos \varphi_{\mathrm{g}j} 
  +\sum_{j<k} 2 \Gamma_j  \Gamma_k \cos\left( \varphi_{\mathrm{g}j} - \varphi_{\mathrm{g}k} \right) 
  \Bigg]   \left| \psi_\mathrm{src} \right|^2
  .  \nonumber 
\end{align}
where we have ignored a small, constant term of order $r_\mathrm{F}/x_j^2$.

As discussed in \citetalias{Gwinn2019}, the geometric phase in this expression, in particular its dependence on $\tau_j$ and $f_j$ in \ref{eq:DefineAlphaBetaGamma}, gives rise to scintillation arcs
in the secondary spectrum. The first term in \ref{eq:AnalyticIntensity} yields a delta-function at the origin;
the second yields a series of delta-functions at positions $(\tau_j, f_j)$  along a parabola, with the behavior expected for the primary scintillation arc;
and the third yields secondary arclets, with apexes along the primary arc and the same curvature, but in the opposite direction.
Traditionally, the brightness of the secondary spectrum is expressed as the square modulus of the intensity in the delay-rate domain,  $C(\tau,f) = | I(\tau, f) |^2$.
In particular, the ratio of the brightness on the primary arc at $(\tau_j, f_j)$ to that at the origin is:
\begin{align}
\frac{C(\tau_j,f_j)}{C(0,0)} &= \frac{\left| I(\tau_j,f_j) \right|^2}{\left| I(0,0) \right|^2} =  \left| \Gamma_j \right|^2 
\end{align}
In this paper, we consider the brightness of an arc as a function of position along the arc, and as a function of time.
This demands a specific model for the widths and the structures of the screen phases within the strips that produce the arc, leading to values of $\Gamma_j$.
From the magnitudes of the screen phases required to reproduce the observed brightness, 
we infer the electron column densities required for the strips.

\begin{figure}
\centering
\includegraphics[width=0.45\textwidth]{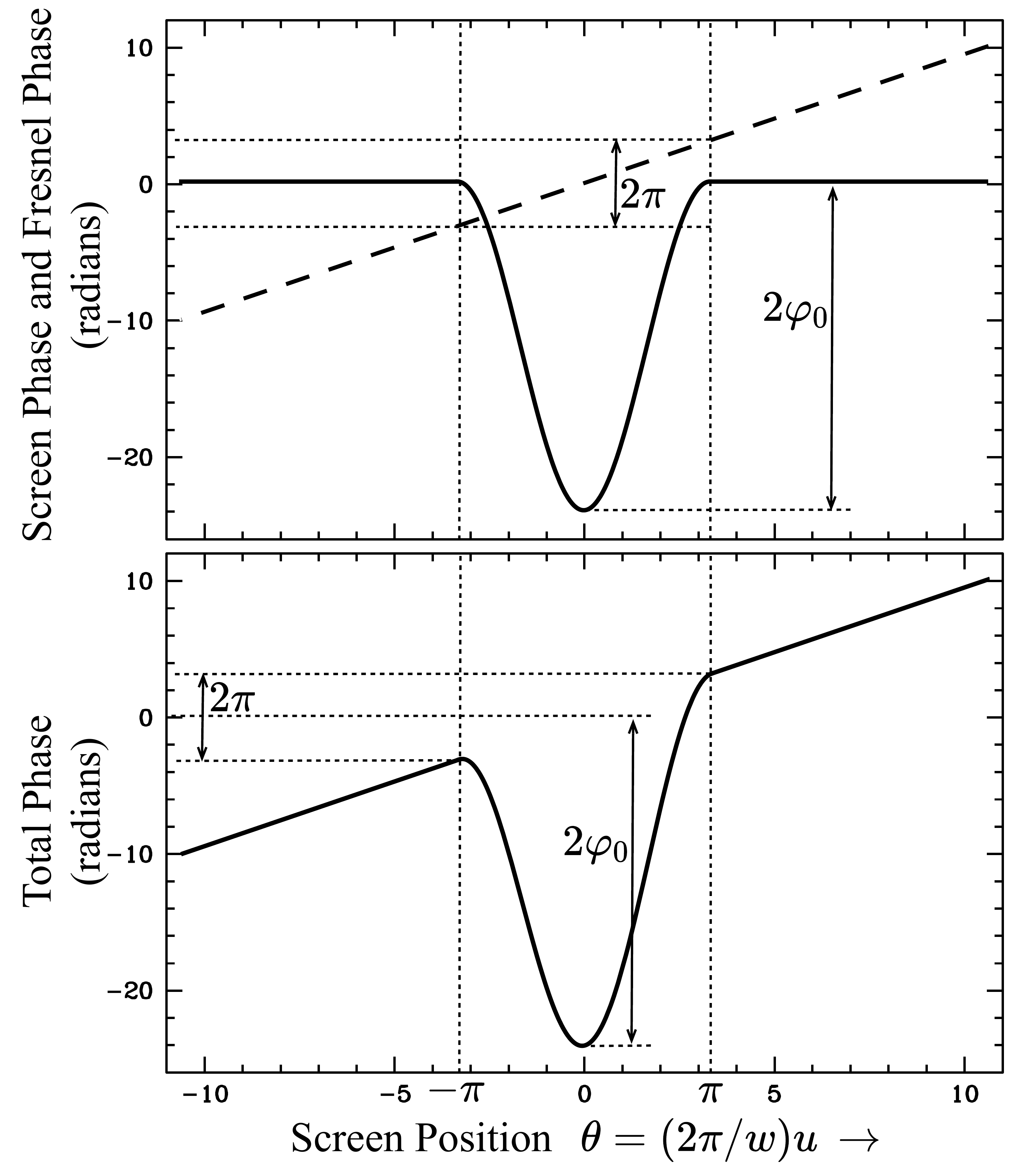} 
\caption{Model phase for analytical noodle model, for $n=1$.  
Dotted vertical lines indicate the edges of the strip.
Upper panel: Solid curve shows the screen phase $\varphi_\mathrm{cos}$ (\ref{eq:CosineModelPhase}),
and dashed curve shows the Fresnel phase.
Lower panel: Solid curve shows the net Fresnel plus screen phase, $\phi_\mathrm{F}(u) + \varphi_\mathrm{cos}(u)$.
\label{fig:IntegrateBlob}}
\end{figure}

\section{Cosine Strip Model}\label{sec:CosineStripModel}

\subsection{Cosine Screen Phase}\label{sec:CosineStripModelIntro}

A simple model offers insight into the optics of strips. 
We suppose that the strip fills $n$ pairs of Fresnel zones on the $x$-axis, so that the Fresnel phase varies by $2\pi n$ across the strip.
Thus, we use\ \ref{eq:W2PiDef} and\ \ref{eq:WidthOfnPairsOfFresnelZones} to set the width of the strip as:
\begin{align}
\label{eq:WStripDef}
w &= n\cdot W_{2\pi} = 2\pi n \frac{r_\mathrm{F}^2}{x_j}
.
\end{align}
We allow $n$ to take on any real value, for greatest generality;
we consider integral and nonintegral values of $n$ separately below.
For our model, 
we consider a screen phase that varies as a cosine wave within the strip, offset so that the strip phase reaches zero at 
either edge of the strip:
\begin{align}
\label{eq:CosineModelPhase}
\varphi_\mathrm{cos}(u) &= - \varphi_0 \left( 1+ \cos\left( \frac{2\pi}{w}  u \right)\right),\quad\quad  - \pi  < \left( \frac{2\pi}{w}  u  \right) < \pi 
.
\end{align}
Again $u=x-x_j$ is the coordinate from the center of the strip, and the edges lie at $u=\pm w/2$.
Figure\ \ref{fig:IntegrateBlob} shows the phase as a function of $u$, for the important special case of $n=1$.
The model has the properties that the screen phase is continuous and singly-differentiable everywhere: it is smooth. 

Equation \ref{eq:DiffractiveStripSimplified} gives, for the contribution to the observed field of the model strip:
\begin{align}
\frac{\psi_{nj} }{ \psi_\mathrm{src}  }
&= \sqrt{ \frac{-i }{2\pi} } \frac{1}{r_\mathrm{F}} e^{ i \phi_{\mathrm{g}j}}     \int_{-w/2}^{w/2} du\, e^{ i  \frac{x_j}{r_\mathrm{F}^2} u} e^{ - i \varphi_0 \left(1 +  \cos \left( \frac{2\pi}{w} u \right) \right)  } 
\nonumber \\
\label{eq:DiffractiveStripAnalyticalModel}
&= e^{ i \phi_{\mathrm{g}j}}    \left(\frac{w}{r_\mathrm{F}}\right)  S_{nj} 
,
\end{align}
where
\begin{align}
\label{eq:SjDef}
S_{nj} &=  \sqrt{ \frac{-i }{(2\pi)^3} }   \int_{-\pi}^{+\pi} d\theta\, e^{ i  n \theta } e^{ - i \varphi_0 (1 +  \cos \theta )  } 
,
\end{align} 
and where we have changed the variable of integration to 
\begin{align}
\label{eq:ThetaDef}
\theta= \frac{2\pi}{w} u,
\end{align}
and used the fact that
\begin{align}
\label{eq:NAltDef}
n &= \frac{w}{2\pi} \frac{x_j}{r_\mathrm{F}^2}
.
\end{align}
The contribution of the strip with zero screen phase, $\psi_{nj0}$, takes precisely the form analogous to\ \ref{eq:DiffractiveStripAnalyticalModel}\ and\ \ref{eq:SjDef}, but with $\varphi_\mathrm{s}=0$:
\begin{align}	
\label{eq:DiffractiveStripAnalyticalModelNumerical_0ScreenPhase}
\frac{ \psi_{nj0} }{  \psi_\mathrm{src} }
&= e^{ i \phi_{\mathrm{g}j}}    \left(\frac{w}{r_\mathrm{F}}\right)   S_{nj0} 
,
\end{align}
where
\begin{align}
S_{nj0} &=  \sqrt{ \frac{-i }{(2\pi)^3} }    \int_{-\pi}^{+\pi} d\theta\, e^{ i  n \theta }  
\nonumber \\
\label{eq:Sj0Def}
& =  \sqrt{ \frac{-i }{(2\pi)^3} }   \frac{2}{n} \sin(n \pi),\quad\mathrm{for\ } n\ne 0
.
\end{align}
Note that $S_{nj}$ and $S_{nj0}$ depend on $x_j$ and $w$, through its dependence on $n$.
We then find:
\begin{align}
\label{eq:Gammaj_from_SnjSnj0}
\Gamma_{j} &=\frac{\psi_{\mathrm{s}j} -\psi_{\mathrm{s}j 0} } { \psi_\mathrm{src} } 
=    \left(\frac{w}{r_\mathrm{F} }\right) \left( S_{nj} - S_{nj0}\right) \\
\label{eq:Gammaj2_from_SnjSnj0}
\left| \Gamma_{j} \right|^2 &=   \left(\frac{w}{r_\mathrm{F} }\right)^2 \left| S_{nj} - S_{nj0}\right|^2
\end{align}

The cosine model as described by\ \ref{eq:SjDef},\ \ref{eq:Sj0Def}, \ref{eq:Gammaj_from_SnjSnj0}, and \ref{eq:Gammaj2_from_SnjSnj0} has three parameters: the strip amplitude $\varphi_0$, its width $w$,
and the number of Fresnel zones it spans $n$. Because the width $w$ and the offset of the strip $x_j$ determine the spacing of Fresnel zones, $n$ can be expressed in terms of $w$ and $x_j$, as \ref{eq:NAltDef} states.
Thus, we choose for model parameters the strip amplitude $\varphi_0$ and any two of the three $w$, $n$, and $x_j$.
In the rest of the paper, we will make use of different sets of these parameters.
We will continue with the set $\left\{ \varphi_0, w, n\right\}$ through the rest of Section\ \ref{sec:CosineStripModel},
as we consider various ways of computing $S_j - S_{j0}$ and so $\Gamma_j$.
To describe the time evolution of $\Gamma_j$ in Section\ \ref{sec:TimeEvolution}, we will use the same set $\left\{ \varphi_0, w, n \right\}$.
To compare $\Gamma_j$ for strips with different $w$ and $\varphi_0$ at a given offset $x_j$ in Sections\ \ref{sec:BrightnessAtOneXj} and\ \ref{sec:NeighboringStrips}, we adopt the set $\left\{ \varphi_0, n, x_j \right\}$.
For comparison of observations with theory in Section\ \ref{sec:Observations}, 
we use primarily $\left\{ \varphi_0, n, x_j \right\}$.

\subsection{Analytical Model}\label{sec:AnalyticalModel}

\begin{figure}
\centering
\includegraphics[width=0.49\textwidth]{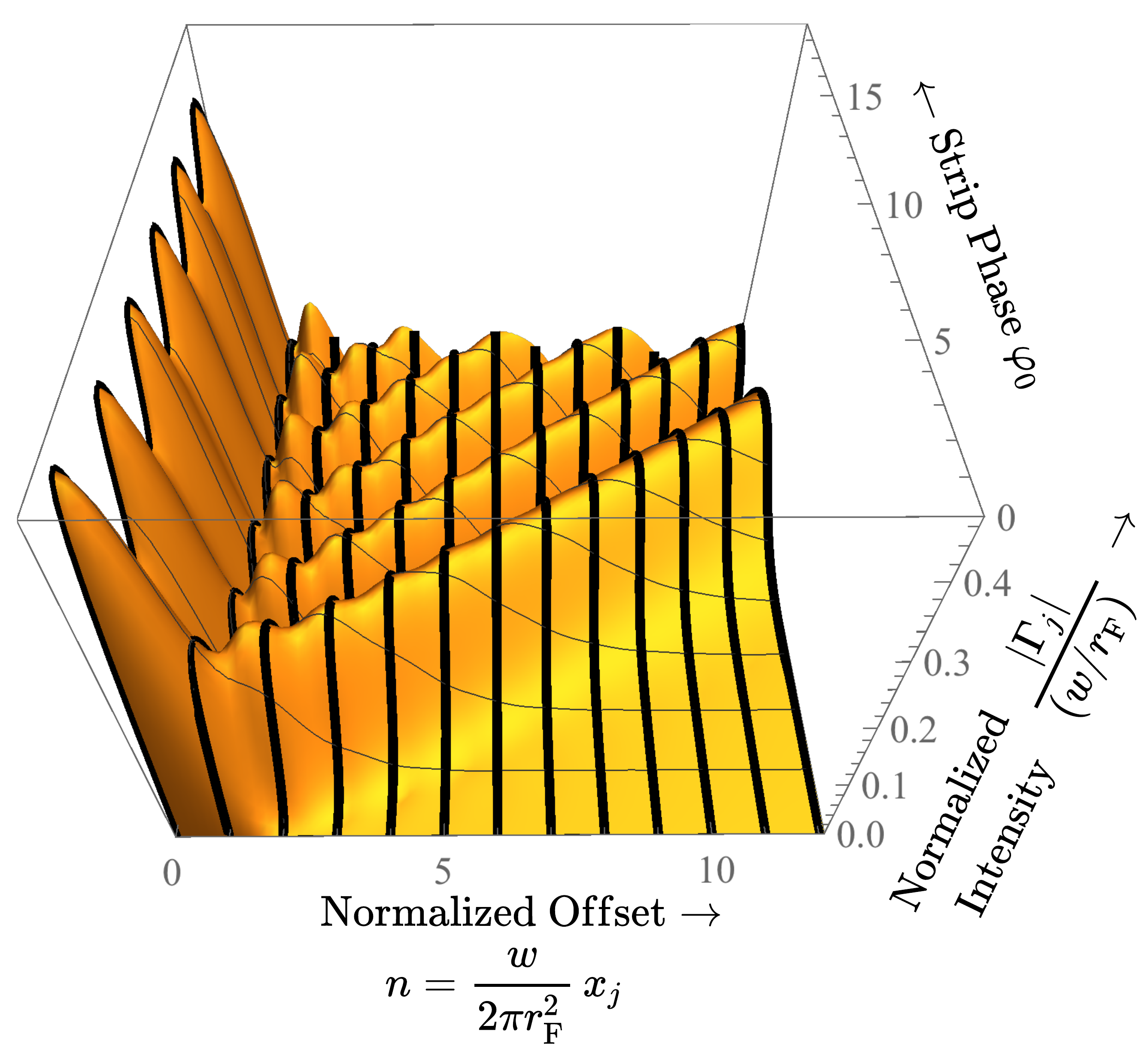} 
\caption{Normalized intensity of a strip, $|\Gamma_{j}|/(w/r_\mathrm{F}) = |S_{nj} - S_{nj0}|$,
plotted as a function of the phase amplitude of the strip $\varphi_0$, and the number of pairs of Fresnel zones spanned by the width of the strip, $n$.  
The normalized intensity is given by\ \ref{eq:CosineModelPhase},\ \ref{eq:SjDef},\ \ref{eq:Sj0Def}, and\ \ref{eq:Gammaj_from_SnjSnj0}.
Recall that
brightness is the square of intensity.
Thick black curves indicate analytical results
as given by\ \ref{eq:SjAnalytical},\ \ref{eq:Sj0Analytical},\ \ref{eq:Snj=0},\ \ref{eq:Snj0=0}, and\ \ref{eq:Gammaj_from_SnjSnj0}.
\label{fig:SameViewPanel1_Vary_Xj}}
\end{figure}

If the strip width $w$ spans an integral number of pairs of Fresnel zones, $n$ is an integer. In this case,
an analytic expression gives the field at the observer.
For this Section\ \ref{sec:AnalyticalModel} only, we specialize to this case.
We find that for nonzero integral $n$ ($n \in \mathbb{Z}_{\ne 0}$):
\begin{align}
\int_{-\pi}^{+\pi} d\theta\, e^{ i  n \theta } e^{ - i \varphi_0 (1 +  \cos \theta )  } 
&=    2\pi i^{-n} e^{-i \varphi_0} J_n(\varphi_0) 
,
\end{align}
where $J_n$ is the familiar regular Bessel function of the first kind, of order $n$.
Note that $J_n$ is identical for positive and negative $n$.
Therefore, we find:
\begin{align}
\label{eq:SjAnalytical}
S_{nj}
&= \sqrt{ \frac{-i }{2\pi } }  \left[ i^{-n} e^{-i \varphi_0} J_n(\varphi_0) \right]\quad\mathrm{for\ }n \in \mathbb{Z}_{\ne 0}
.
\end{align}
Because $n$ is a nonzero integer,
\ref{eq:Sj0Def} shows that the contribution of the strip with zero screen phase vanishes:
\begin{align}
\label{eq:Sj0Analytical}
S_{nj 0}&= 0 \quad\mathrm{for\ }n \in \mathbb{Z}_{\ne 0}
.
\end{align}
From \ref{eq:Gammaj_from_SnjSnj0}, we find for the contribution to the field, and the brightness of the resulting arc:
\begin{align}
\label{eq:AnalyticalModelExactBesselFunctionN}
\Gamma_{j} &
=\frac{i^{-n-\frac{1}{2}}}{\sqrt{2\pi }}  \left( \frac{w}{r_\mathrm{F}} \right)  e^{-i \varphi_0} J_n(\varphi_0) 
 \quad\mathrm{for\ }n \in \mathbb{Z}_{\ne 0}\\
\left| \Gamma_{j} \right|^2 &= \frac{1}{2 \pi} \left( \frac{ w}{r_\mathrm{F}}\right)^2 \left| J_n(\varphi_0) \right|^2
 \quad\mathrm{for\ }n \in \mathbb{Z}_{\ne 0}
.
\end{align}
The intensity of a point on the arc is proportional to $|\Gamma_j|$, as\ \ref{eq:AnalyticIntensity} states.
Figure\ \ref{fig:SameViewPanel1_Vary_Xj} displays the normalized intensity $|\Gamma_j|/(w/r_\mathrm{F}$, from\ \ref{eq:AnalyticalModelExactBesselFunctionN}.
The small factor of $w/ r_\mathrm{F}$ means that the magnitude of $\Gamma_{j}$ will be small relative to 1,
and so the intensity of the point will be less than that of the undeflected path.
Indeed, the contribution of the strip to the field at the observer has an effective area of $\sqrt{2\pi} w r_\mathrm{F}$, whereas that of the undeflected line of sight has effective area $2\pi r_\mathrm{F}^2$, as comparison of\ \ref{eq:DiffractiveStripAnalyticalModel}-\ref{eq:SjDef} and \ref{eq:DiffractiveStripAnalyticalModelNumerical_0ScreenPhase}-\ref{eq:Sj0Def} with 
Equations\ 4 
and\  11 in \citetalias{Gwinn2019} 
shows.

Very narrow strips span much less than a pair of Fresnel zones, so that $n\rightarrow 0$. 
We calculate the contribution $\psi_n$ for $n= 0$ using L'Hospital's rule,
and use the fact that $\psi_n$ is continuous in $n$ to gain insight into the scattering of very narrow strips.
We find for $n=0$:
\begin{align}
\label{eq:Snj=0}
S_{0j}
&=   \sqrt{ \frac{-i }{2\pi } }   \left[ e^{-i \varphi_0} J_0(\varphi_0) \right]
\\
\label{eq:Snj0=0}
S_{0j0} &= \sqrt{ \frac{-i }{2\pi } }
.
\end{align}
The last expression shows that the contribution for a strip with no screen is nonzero for $n=0$, unlike the case for other integral values of $n$ in\ \ref{eq:Sj0Analytical}.
This offset is significant: it gives rise to a ``wall'' of intensity at $n=0$ in parameter space, as discussed in the following section.
Thus, for a very narrow strip that spans much less than a pair of Fresnel zones, so that $|n|\ll 1$, the contribution to the field and brightness are:
\begin{align}
\label{eq:AnalyticalModelExactBesselFunction0}
\Gamma_{0j} 
&= \sqrt{ \frac{-i }{2\pi} } \left( \frac{w}{r_\mathrm{F}} \right) \left( e^{-i \varphi_0}  J_0(\varphi_0) -1 \right)
 \quad\mathrm{for\ }n=0
\\
\label{eq:Gamma2AnalyticalModelExact0}
\left| \Gamma_{0j} \right|^2 &= \frac{1}{2 \pi} \left( \frac{ w}{r_\mathrm{F}}\right)^2 \left| J_0(\varphi_0) -1 \right|^2
 \quad\mathrm{for\ }n=0
.
\end{align}

\subsection{Discussion: Model Intensity}\label{sec:NumericalExtensionAndDiscussion}

Figure\ \ref{fig:SameViewPanel1_Vary_Xj} displays the normalized intensity $\Gamma_j/(w/r_\mathrm{F})$,
as a function of $n$ and $\varphi_0$ for $n>0$.
Although the field at the observer depends on three parameters $\left\{ \varphi_0, w, x_j\right\}$, the integrals $S_{nj}$ and $S_{nj0}$ depend on only two: $\varphi_0$ and $n$.
Thus, we display $|S_{nj} - S_{nj0}| = |\Gamma_j|/(w/r_\mathrm{F})$
as a function of $\varphi_0$ and $n$ only.
For non-integral values of $n$, we must numerically evaluate\ \ref{eq:SjDef} and\ \ref{eq:DiffractiveStripAnalyticalModelNumerical_0ScreenPhase} through\ \ref{eq:Gammaj2_from_SnjSnj0}. Figure\ \ref{fig:SameViewPanel1_Vary_Xj} shows the function as a surface.
The analytical results, for integer $n$, form a skeleton for the function.

To help understand the plot, we may suppose that the width of the strip $w$ is held constant, and we vary $\varphi_0$ and $n$.
For constant width $w$, a varying $n=(w/2\pi r_\mathrm{F}^2) x_j$ plays the role of a normalized offset $x_j$:
change of $n$ corresponds to displacement of the strip toward or away from the optical axis.
Negative $n$ is possible for $x_j<0$, but $|\Gamma_j|$ depends only on the magnitude of $n$, so we may imagine this plot as the positive side of a plot symmetric about $n=0$.  

The overall features of the plot of $ |\Gamma_j|/(w/r_\mathrm{F})$ are: 
a series of parallel ``ridges'' in the triangular region with $\varphi_0 > n$;
a ``flat'' triangular region with nearly 0 intensity, in the triangle $\varphi_0 < n$;
and a narrow ``wall'' with crenellations for $n\rightarrow 0$.
Properties of Bessel functions prescribe the form of this plot.
The Bessel function $J_n(\varphi_0)$ reaches an absolute maximum at $\varphi_0 \approx n+1$. 
This is the first and highest of the series of ridges in the triangle $\varphi_0 > n$.
The magnitude of that maximum declines approximately as $n^{-1/3}$, so the ridge is nearly flat.
Subsequent local maxima define the ridges.
At smaller values $\varphi_0$, the Bessel function increases only slowly, as $J_n(\varphi_0) \propto \varphi_0^n$, from 0 at $\varphi_0=0$;
this defines the flat region with nearly 0 intensity.
For $n\rightarrow 0$, recall from \ref{eq:Snj0=0} that the contribution of the screen with zero phase is a nonzero constant (\ref{eq:Snj0=0});
this gives rise to the wall. The linear rise from 0 at $\varphi_0=0$ and the crenellations along the wall arise from partial cancellation of that constant by the contribution of the strip \ref{eq:Snj=0}.

\subsection{Small Electron Column}\label{sec:SmallElectronColumn}

\begin{figure}
\centering
\includegraphics[width=0.48\textwidth]{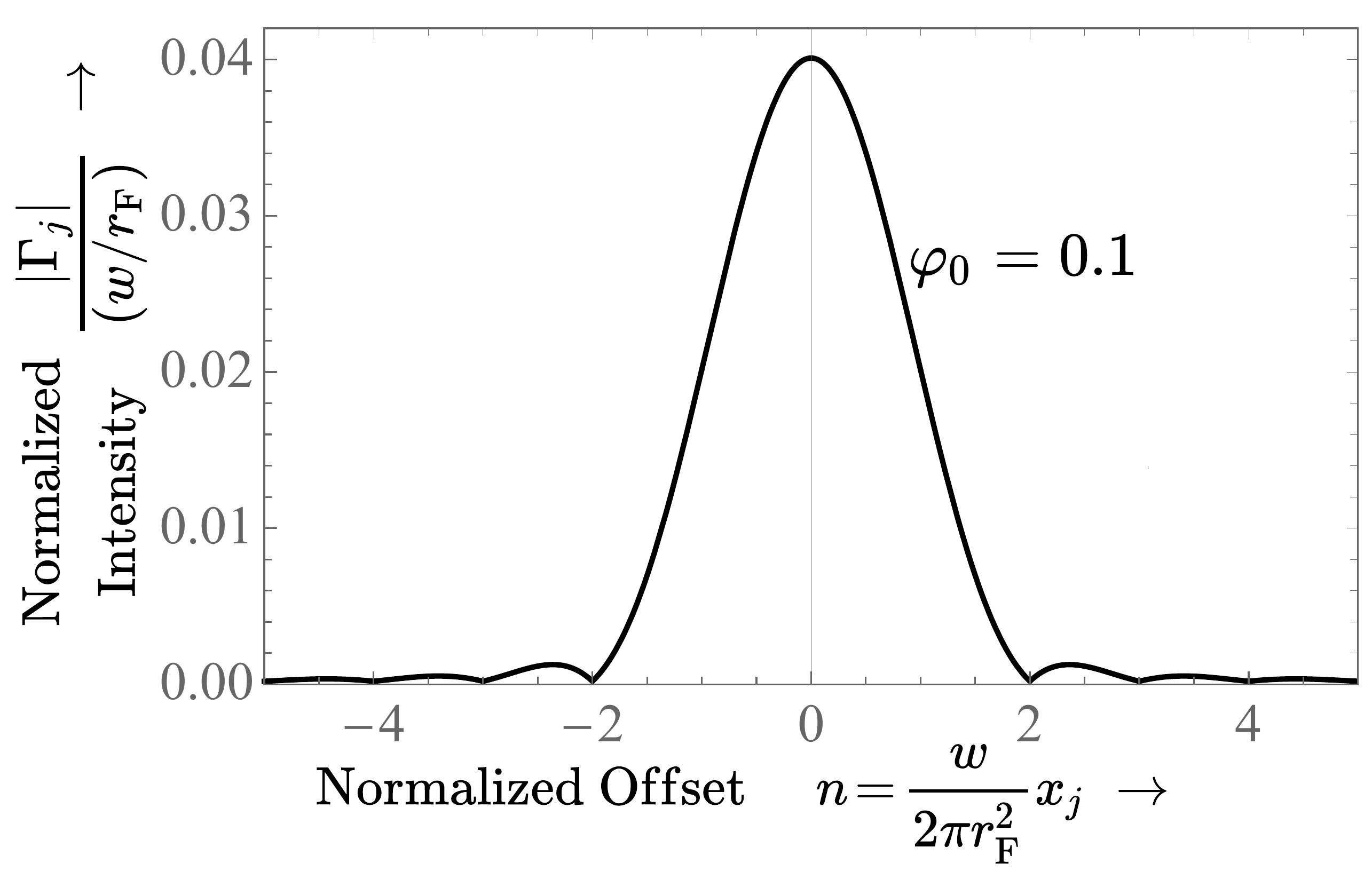} 
\caption{Normalized intensity $|{\Gamma_{j}}|/({w/r_\mathrm{F}})$,
plotted with normalized offset $n$ for $\varphi_0=0.1$.
The approximate form from\ \ref{eq:IntegralPhi0SmallAnalyticForm} is indistinguishable from the exact form on this plot.
The same form holds for smaller $\varphi_0$, with a proportionate reduction of the normalized intensity.
Note that $n=w/W_{2\pi}$.
\label{fig:SmallColumn}}
\end{figure}

We now find an expression for the contribution to the field $|\Gamma_{j}|$ from strips with low electron columns, so that $\varphi_0 < 1$.
As Figure\ \ref{fig:SameViewPanel1_Vary_Xj} shows, $|S_{nj} - S_{nj0}|=0$ at $\varphi_0 =0$ for all $n$,
so we consider the first-order approximation
\begin{align}
\label{eq:IntegralForSmallVarFi0}
\frac{\Gamma_{j}}{w/r_\mathrm{F}} = S_{nj} - S_{nj0} & \approx \varphi_0 \left[ \frac{\partial}{\partial \varphi_0} \left( S_{nj} - S_{nj0}\right)\right]_{\varphi_0=0} 
\end{align}
We differentiate $S_{nj}$ (as given by \ref{eq:SjDef}) by $\varphi_0$, and note that $S_{nj0}$ is independent of $\varphi_0$.
We find:
\begin{align}
\frac{\Gamma_j}{w/r_\mathrm{F}} &= S_{nj} - S_{nj0} 
\nonumber \\
\label{eq:IntegralPhi0SmallAnalyticForm}
& \approx \sqrt{\frac{-i}{(2\pi)^3}} \int_{-\pi}^{\pi} d\theta\; \varphi_0\cdot \left[ \frac{\partial}{\partial \varphi_0}e^{i n \theta - \varphi_0(1+\cos \theta)} \right]_{\varphi_0=0} \\
\label{eq:ResultPhi0SmallAnalyticForm}
& =\sqrt{\frac{-i}{(2\pi)^3}} \frac{2 i \sin(n \pi)}{n^3-n} \varphi_0 
.
\end{align}
Figure\ \ref{fig:SmallColumn} shows this function for $\varphi_0=0.1$; the approximation\ \ref{eq:IntegralPhi0SmallAnalyticForm} is indistinguishable from the exact value,
at the scale of the plot. 
For smaller values of $\varphi_0$, the approximation is even better and ${\Gamma_j}/{(w/r_\mathrm{F})}$ takes the same form, scaled by $\varphi_0$.
As the figure shows, a strip with small electron column will produce a significant intensity for $|n| < 2$,
but otherwise produces only a very small one.
Of course, this corresponds to the fact that a strip scatters most effectively when its width is about one pair of Fresnel zones, as noted above and shown as the ``wall'' in Figure\ \ref{fig:SameViewPanel1_Vary_Xj}.
Variation of Fresnel phase across the strip tends to cancel out the contributions of wider strips. 

For $n\rightarrow 0$, with L'Hospital's rule the approximation\ \ref{eq:ResultPhi0SmallAnalyticForm} yields:
\begin{align}
\frac{\Gamma_j}{w/r_\mathrm{F}} 
&= S_{0j} - S_{0j0}
\approx\sqrt{\frac{-i}{(2\pi)^3}} (- i 2 \pi \varphi_ 0)\nonumber \\
\label{eq:IntegralPhi0SmallAnalyticForm_n=1}
\frac{\left|\Gamma_j\right|}{w/r_\mathrm{F}} 
&= \frac{1}{\sqrt{2 \pi }}\, \varphi_0
.
\end{align}
For $n\rightarrow 1$, again with L'Hospital's rule, the approximation yields:
\begin{align}
\frac{\Gamma_j}{w/r_\mathrm{F}} 
&= S_{1j} - S_{1j0}
\approx\sqrt{\frac{-i}{(2\pi)^3}} (- i \pi \varphi_ 0) \nonumber \\
\label{eq:IntegralPhi0SmallAnalyticForm_n=1}
\frac{\left|\Gamma_j\right|}{w/r_\mathrm{F}} 
&= \sqrt{\frac{1}{8 \pi}} \varphi_0 
.
\end{align}
The approximation for the amplitude is good to about 14\% for $\varphi_0=1$. The approximation is much poorer for the complex phase.

\subsection{Stationary-Phase Approximation}\label{sec:StatPhaseApprox}

Our simple cosine model provides an illuminating example of the stationary-phase approximation \citep[see, for example,][]{1978amms.book.....B}.
In this approximation, an integral of a complex function with rapidly-varying phase and slowly-varying amplitude 
is represented as a sum, over the points where the phase is constant as a function of the integrand;
in other words, where the derivative of the phase vanishes.
These points are known as points of stationary phase.
The sum is weighted by the inverse of the curvature of the phase at those points;
or more precisely, by the reciprocal of the square root of the absolute value of the second derivative of phase.
Thus, in this calculation, each point of stationary phase is reckoned to contribute over a length about equal to that of the region where the phase remains within $\pi/4$ radians of its value at the point of stationary phase.
This length, times the length $\sqrt{2\pi} r_\mathrm{F}$ in the $y$-direction, equals the effective area for the un-normalized Kirchhoff integral,
as seen from comparison with\ \ref{eq:KirchhoffIntegral} and\ \ref{eq:FieldFromStrip}.
Outside of that region, the phase rotates rapidly and contributions to the integral cancel out.
The approximation obviously requires that points of stationary phase exist,
and that they be separated by far enough that the rotating residual phases can cancel out.
The stationary-phase approximation is closely related to the ray approximation;
indeed,
rays are the paths of stationary phase difference between source and observer, relative to all nearby paths.
Consequently, they are extremal in travel time, at the phase velocity along the path.
(They might also lie at saddle points of travel time.)
Here, we investigate the stationary-phase approximation in the context of our model.

\begin{figure}
\centering
\includegraphics[width=0.45\textwidth]{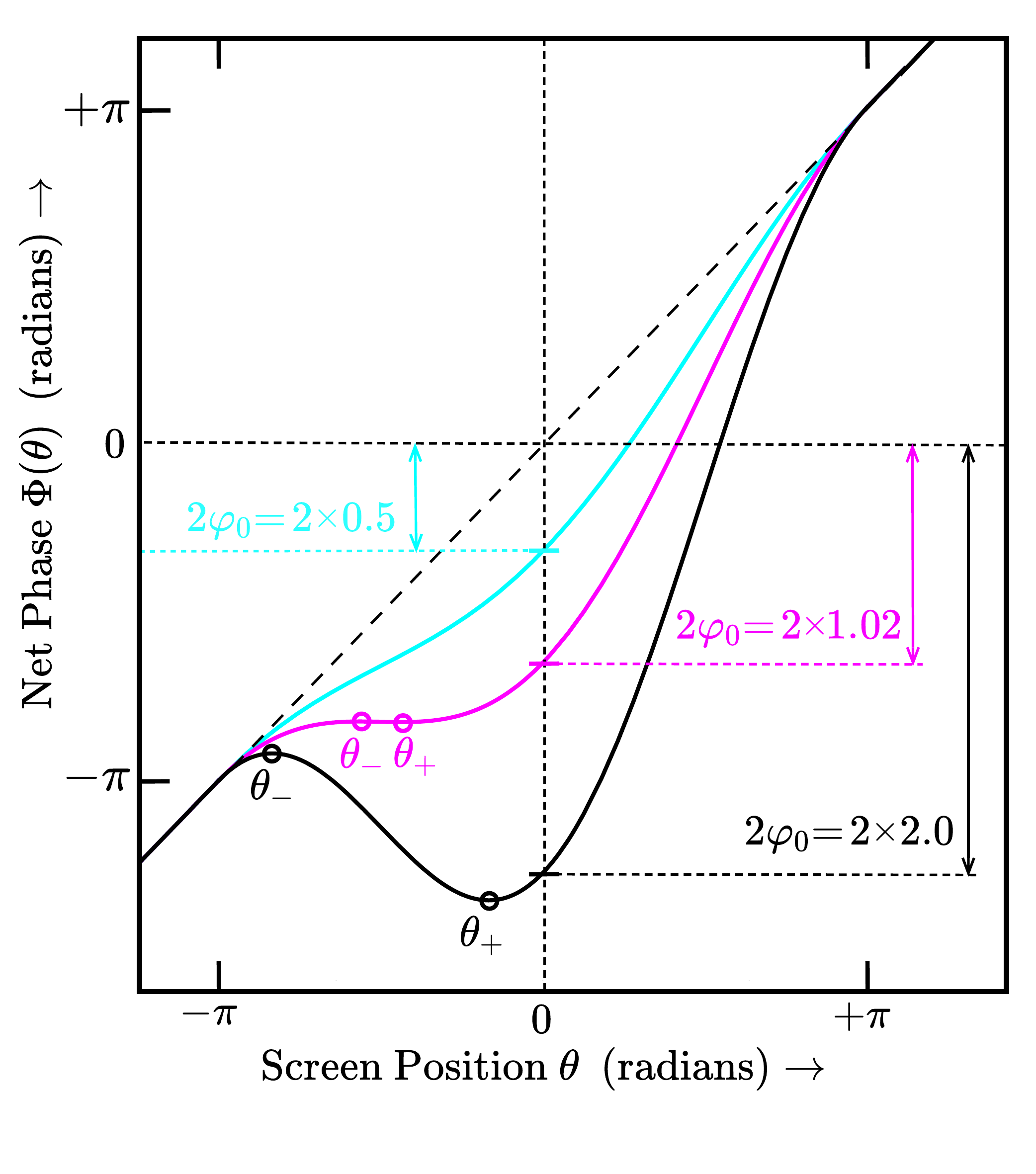} 
\caption{Cosine models for stationary phase approximation in the case $n=1$, for amplitudes of screen phase $\varphi_0 =0.5, 1.02$, and $2.0$ radians (cyan, magenta, and black curves). Stationary-phase points $\theta_-$ and $\theta_+$ are shown for cases $\varphi_0 =1.02$ and $2.0$;
for $\varphi_0 =0.5$ no stationary-phase points exist.
\label{fig:StatPhasePoints}}
\end{figure}

For convenience, we define for the net phase of the integrand: 
\begin{align}
\Phi(\theta) = n \theta - \varphi_0(1+\cos\theta)
,
\end{align}
thus combining the local Fresnel and screen phases into $\Phi(\theta)$. 
We seek the integral:
\begin{align}
\sigma_{nj} &= \int_{-\pi}^\pi d\theta\, e^{i\Phi(\theta)}
.
\end{align}
The contribution of the strip to the field is given by:
\begin{align}
S_{nj} &=  \sqrt{ \frac{-i }{(2\pi)^3} }    \sigma_{nj} 
\end{align}
along with\ \ref{eq:Sj0Def} and \ref{eq:Gammaj_from_SnjSnj0}.
The area corresponding to the un-normalized Kirchhoff integral is $\sqrt{2\pi} r_\mathrm{F} | \sigma_{nj}|$.
For our discussion of the stationary-phase approximation, we suppose $\varphi_0>0$ for simplicity,
so that the screen phase $\varphi_\mathrm{cos} < 0$, as shown in Figures\ \ref{fig:IntegrateBlob} and\ \ref{fig:StatPhasePoints}.
Extension to $\varphi_0<0$ is straightforward.

For large values of $\varphi_0$, the net phase $\Phi(\theta)$ has two points of stationary phase, as Figure\ \ref{fig:StatPhasePoints} shows.
One is concave downward and lies at $\theta_-$ in the range $-\pi <\theta\le -\pi/2$. 
The other is concave upward and lies at $\theta_+$ in the range $-\pi/2 \le \theta<0$. 
As $\varphi_0 \rightarrow \infty$, the points move toward the extremes of their ranges, at $-\pi$ and $0$, respectively. At $\varphi_0=n$, the two points merge at $\theta=\pi/2$. At still smaller values, of $\varphi_0 < n$, no stationary phase points exist. 
The corresponding integral for the strips without screen phase, $S_{j0}$, has no points of stationary phase,
and does not contribute in the stationary-phase approximation.

Stationary-phase points lie where $\frac{\partial }{\partial \theta} \Phi(\theta) =0$, or for our model at:
\begin{align}
\theta_{-} &=- \pi + \arcsin\left(n/\varphi_0\right)  \\                   
\theta_{+} &= \phantom{\pi - }- \arcsin \left(n/\varphi_0 \right)  
\end{align}
The integrand $e^{i\Phi (\theta)}$ takes on the values at those points:
\begin{align}
e^{i \Phi(\theta_{-}) }&=                  e^{-i n \pi}    e^{i\left(- \varphi_0 +\sqrt{\varphi_0^2-n^2}  + n \arcsin \left( n/\varphi_0 \right) \right) } \\
e^{i\Phi(\theta_{+}) }&= \phantom{ e^{-i n \pi} } e^{i \left(- \varphi_0 - \sqrt{\varphi_0^2-n^2}  - n \arcsin \left( n/\varphi_0 \right) \right) }
\end{align}
The second derivatives of $\Phi$ at those points are:
\begin{align}
\left[ \frac{\partial^2  \Phi}{\partial^2 \theta}\right]_{\theta_{-}}&= - \sqrt{\varphi_0^2 -n^2} \\
\left[ \frac{\partial^2  \Phi}{\partial^2 \theta} \right]_{\theta_{+}}&= \phantom{-} \sqrt{\varphi_0^2 -n^2}
\end{align}
Consequently, the approximation yields for the integral:
\begin{align}
&\sigma_{nj}
\approx
\sqrt{\frac{2 \pi}{\left| \frac{\partial^2  \Phi}{\partial^2 \theta} \right|_{\theta_{-}}}}
 e^{i \left(\Phi (\theta_{-})  - \frac{\pi}{4} \right)}
+
 \sqrt{\frac{2 \pi}{\left| \frac{\partial^2  \Phi}{\partial^2 \theta}  \right|_{\theta_{+}}}}
e^{i \left(\Phi(\theta_{+})  + \frac{\pi}{4} \right)}
\nonumber
 \\
\label{eq:StatPhaseJnActual}
 &\quad \approx \frac{\sqrt{8\pi} }{\sqrt[4]{\varphi_0^2-n^2}} i^{-n}e^{- i\varphi_0}\cos \left(\sqrt{\varphi_0^2-n^2} + \arcsin \left(\frac{n}{\varphi_0} \right) 
 \small{ -\frac{\pi }{2}(n+\textstyle{\frac{1}{2}})  }\right)
  .
\end{align}
The expression \ref{eq:StatPhaseJnActual}
is the product of three factors.
The first factor is an amplitude.
It declines with increasing $\varphi_0$, as curvature at the stationary-phase points sharpens and the characteristic length of the region with total phase $\Phi \lesssim \pi/4$ decreases. It diverges as $\varphi_0$ decreases to $n$,
as the curvature decreases, the characteristic length increases, and the stationary-phase points converge and merge.
The second factor is a phase, increasing linearly with $\varphi_0$.
It is the average of the complex phases at the stationary-phase points. 
The third factor is a real-valued sinusoid; this represents the interference of the two stationary-phase points.

\begin{figure}
\centering
\includegraphics[width=0.45\textwidth]{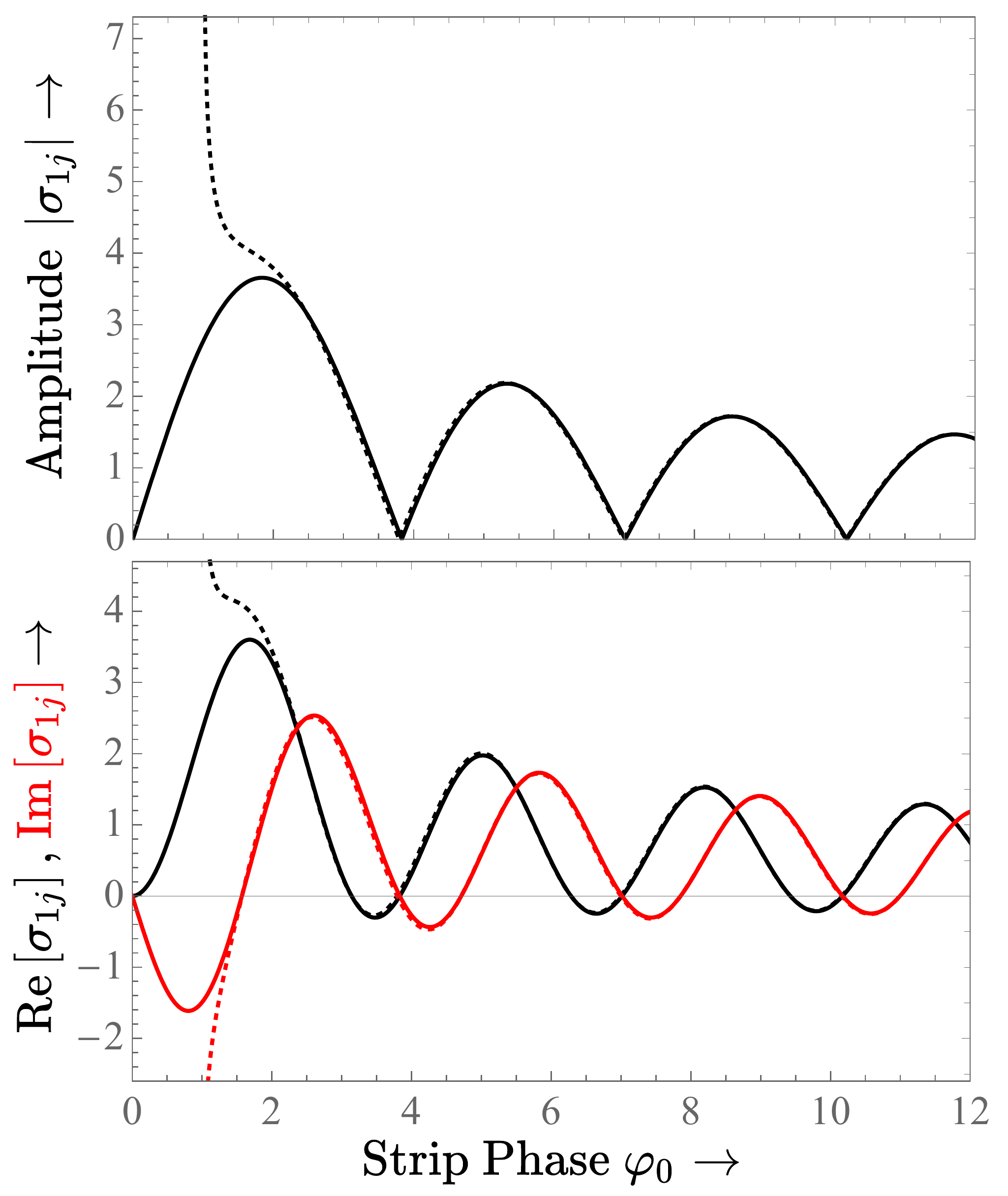} 
\caption{
Comparison of exact values (solid lines:\ \ref{eq:SjAnalytical}) and stationary-phase approximation (dotted lines:\ \ref{eq:StatPhaseJnActual}) for the analytical model of sinusiodal screen phase with $n=1$.
Upper panel: Comparison of amplitude.
Lower panel: Comparison of real part (black) and imaginary part (red).
For $n=1$, the stationary-phase approximation is good for $\varphi_0 \gtrsim 2$.
\label{fig:StatPhasePlot}}
\end{figure}

Comparison of\ \ref{eq:StatPhaseJnActual} with\ \ref{eq:SjAnalytical}, the exact result for integral $n$, shows that the stationary-phase approximation is good for $\varphi_0 \gtrsim n+1$.
As an example, Figure\ \ref{fig:StatPhasePlot} compares the exact and approximate results for $n=1$.
As the plot shows, the approximation is good for $\varphi_0\gtrsim 2$ in this case.
This is perhaps surprising; usually one expects the stationary-phase approximation to hold only when stationary-phase points are separated by many turns of phase in $\Phi(\theta)$ \citep{1978amms.book.....B}. 
For $n=1,\ \varphi_0=2$, they are separated by only $1.37...$ radians.
The approximation fails dramatically for $\varphi_0 <2$, and of course the approximation yields zero for $\varphi_0 < 1$.

\subsubsection{Wave and Ray Regimes}

We divide our noodle model, and such models in general, into two optical regimes: the ray and wave regimes. In the ray regime, the stationary-phase approximation holds, so that the problem can be described as the interference of multiple discrete paths. The only significant paths intersect the screen at the stationary-phase points, where the refracting material locally acts as a prism to deflect radiation to the observer. In the wave regime, on the other hand, the stationary-phase approximation does not hold. Description of the field requires wave theory, and diffraction is important. For our cosine model, the boundary between wave and ray regimes lies at $|\varphi_0|\approx |n|+1$.  
Larger values of $|\varphi_0|$ can be described with ray optics, and smaller values of $|\varphi_0|$ require wave optics.

The well-known ``diffractive'' and ``refractive'' effects of strong scattering are independent categories from the ``wave'' and ``ray'' regimes described here.
Diffractive scattering results from the correlation of phase with position on the scattering screen, and refractive from the correlation of phase gradients \citep{2015ApJ...805..180J}. 
Thus, diffractive and refractive regimens coexist in a single environment \cite{1992RSPTA.341..151N}. 
By contrast, the ``wave'' and ``ray'' regimes, as defined here by the invalidity or validity of the stationary-phase approximation, do not coexist.

\section{Brightness with Time, Width, and Superposition}\label{sec:BrightnessInTimeWidthSuperposition}

\subsection{Time Evolution of Field from a Strip}\label{sec:TimeEvolution}

Over the course of time,
motions of pulsar and observer will move the undeflected line of sight closer to a strip until the line of sight passes through the strip,
and then move it further away.
For multiple epochs of observation, the offset $x_j$ most conveniently describes this motion: it increases linearly with observing epoch,
over days to months.
If the strip is unchanging, then $\varphi_0$ and $w$ remain constant.
Consequently, we adopt the parameter set $\left\{ \varphi_0, w, n \right\}$ and regard $n=(w/2\pi  r_\mathrm{F}^2 ) x_j$ as the ``normalized offset,'' a proxy for $x_j$.
For this parameter set, the intensity $|\Gamma_j |$ depends on $w$ only through the prefactor $(w/r_\mathrm{F})$, as\ \ref{eq:Gammaj_from_SnjSnj0} states.
The ``normalized intensity'' $|\Gamma_j|/(w/r_\mathrm{F})$ depends only on $\varphi_0$ and $n$.
On Figure\ \ref{fig:SameViewPanel1_Vary_Xj},
the strip will move from large $n$ to $n=0$ at constant $\varphi_0$, and then reverse;
alternatively, we imagine it exploring the other half of a plot symmetric about $n=0$.
This motion takes place at constant speed.
For a strip with small electron column $|\varphi_0| \lesssim 1$, the light curve -- the history of intensity with time -- will follow Figure\ \ref{fig:SmallColumn}, with the vertical axis rescaled to the particular value of $\varphi_0$.
On the other hand, for a strip with large electron column $\varphi_0 > 1$, the strip will display $|\Gamma_j|\approx 0$ in the flat region, followed by oscillations as it traverses the ridges, a peak as it crosses the wall, then further oscillations, and finally darkness again.

Brightness in the secondary spectrum is the square of intensity, so the dominance of the wall is even more pronounced for observations of brightness.
This behavior leads to the selection effect discussed in Section\ 3.2.1 
of \citetalias{Gwinn2019}: strips scatter only within a limited distance from the undeflected line of sight, and are most effective at scattering when they cover one pair of Fresnel zones or less.
We discuss observed brightnesses of points on the arc, and their time evolution, more fully in Sections\ \ref{sec:Observations} and\ \ref{sec:Discussion} below.

\subsection{Brightness of Arcs with Width at Constant Offset}\label{sec:BrightnessAtOneXj}

\begin{figure}
\centering
\includegraphics[width=0.49\textwidth]{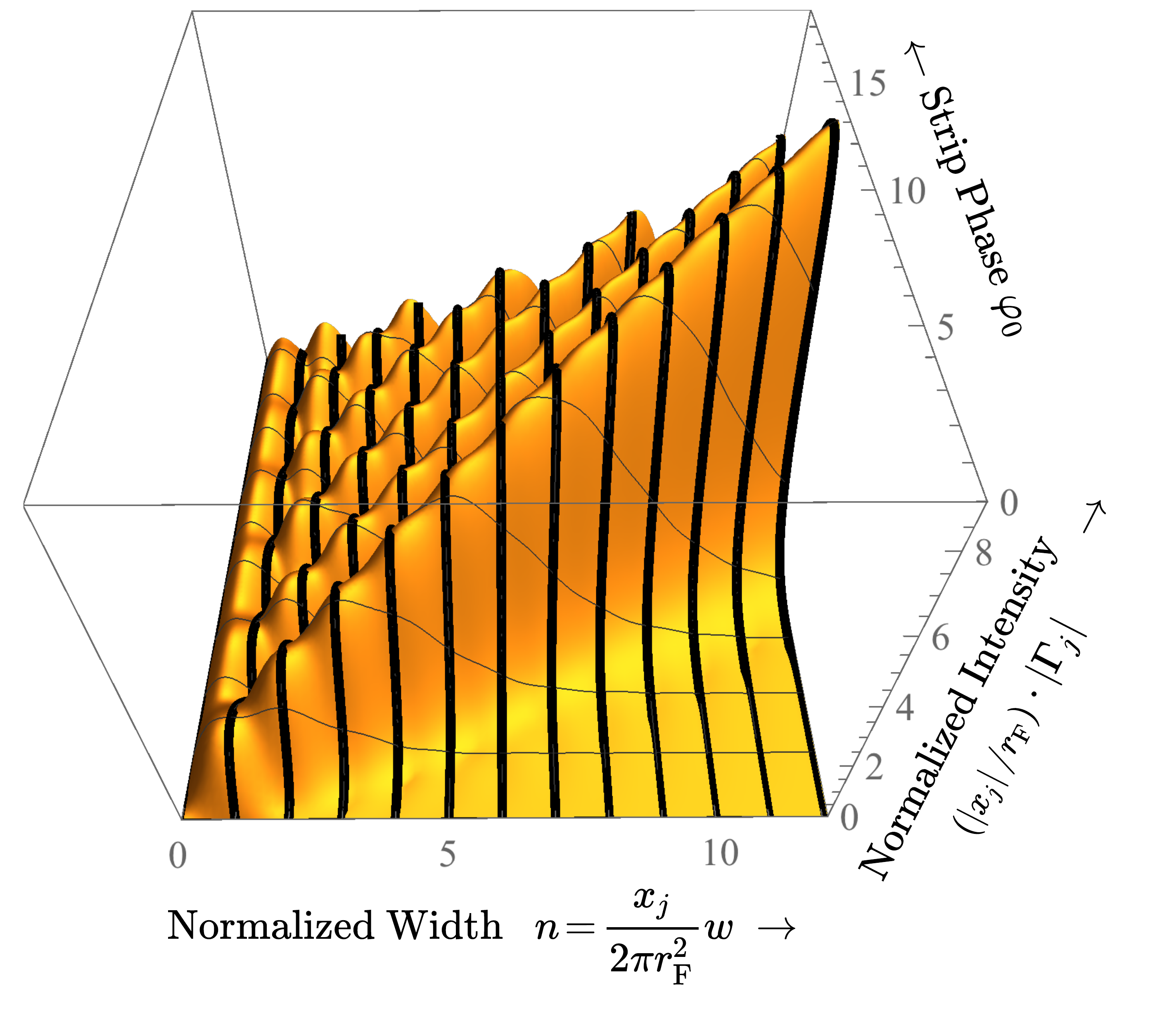}
\caption{Normalized intensity of a strip, $(x_j/r_\mathrm{F}) |\Gamma_{j}| = 2\pi n |S_{nj} - S_{nj0}|$,
plotted as a function of the characteristic phase introduced by the strip $\varphi_0$, and the number of Fresnel zones spanned by the width of the strip $n$.  
The normalized intensity is given by\ \ref{eq:CosineModelPhase},\ \ref{eq:SjDef},\ \ref{eq:Sj0Def}, and\ \ref{eq:Gammaj_from_SnjSnj0_xjForm};
brightness is the square of intensity.
Thick black curves indicate analytical results
as given by\ \ref{eq:SjAnalytical},\ \ref{eq:Sj0Analytical},\ \ref{eq:Snj=0},\ \ref{eq:Snj0=0}, and\ \ref{eq:Gammaj_from_SnjSnj0_xjForm}.
Note that $n=w/W_{2\pi}$.
The perspective is the same as that of Figure\ \ref{fig:SameViewPanel1_Vary_Xj}.
\label{fig:SameViewPanel2_Vary_w}}
\end{figure}

To match observations, we must compare strips of different properties at a particular point in the secondary spectrum, corresponding to a particular offset $x_j$. In this situation, $x_j$ remains constant,
but the phase amplitude $\varphi_0$ and the width $w$ vary.
Consequently, we adopt the parameter set $\left\{ \varphi_0, n, x_j \right\}$ and regard $n=(|x_j|/2\pi r_\mathrm{F}^2) w$ as the ``normalized width,'' a proxy for $w$.
We re-express \ref{eq:Gammaj_from_SnjSnj0} in the form:
\begin{align}
\label{eq:Gammaj_from_SnjSnj0_xjForm}
( |x_j|/r_\mathrm{F}) \, \Gamma_{j} &=2\pi\, n\, \left(S_{nj} - S_{nj0}\right)  
.
\end{align}
The expression on the right depends only on $\varphi_0$ and $n$.
Figure\ \ref{fig:SameViewPanel2_Vary_w} shows the function $(|x_j|/r_\mathrm{F}) \Gamma_{j}$, plotted with $\varphi_0$ and $n$.
This is the same plot as that shown in Figure\ \ref{fig:SameViewPanel1_Vary_Xj}, viewed from the same point in parameter space, but multiplied by $2\pi n$.
Consequently, the wall is absent. 
One can imagine exploring the plot by holding $|x_j|$ constant,
and varying $\varphi_0$, and $n$ as a proxy for $w$, to find the intensities of different strips at a given offset $x_j$.

The analytical results provide a skeleton for Figure\ \ref{fig:SameViewPanel2_Vary_w}.
For integral $n$, \ref{eq:Gammaj_from_SnjSnj0_xjForm} takes the form given by using \ref{eq:AnalyticalModelExactBesselFunctionN}.
The normalized intensity is:
\begin{align}
\label{eq:Gammaj_from_SnjSnj0_xjFormInteger}
(|x_j|/r_\mathrm{F})
&=i^{-n-\frac{1}{2}} \sqrt{2\pi } n  e^{-i \varphi_0} J_n(\varphi_0)
\quad\mathrm{for\ }n \in \mathbb{Z}_{\ne 0}
.
\end{align}
Similarly, the normalized brightness is:
\begin{align}
\label{eq:Gammaj2_from_SnjSnj0_xjFormInteger}
(|x_j|/r_\mathrm{F})^2 \left|\Gamma_{j}\right|^2 
&=2\pi n^2  J_n(\varphi_0)^2
\quad\mathrm{for\ }n \in \mathbb{Z}_{\ne 0}
.
\end{align}
We are particularly interested in the case of $\varphi_0 \ll 1 $. Analogously to \ref{eq:ResultPhi0SmallAnalyticForm}, we find for the case of varying $w$ and constant $x_j$:
 \begin{align}
\label{eq:OffsetSlopePhi0SmallAnalyticFixXj}
(|x_j|/r_\mathrm{F}) \Gamma_{j} 
&= \sqrt{\frac{2 i}{\pi}} \frac{\sin(n \pi)}{n^2-1}  \varphi_0 
\quad\mathrm{for\ }\varphi_0 \ll 1
.
\end{align}
This has the interesting limits for $n\rightarrow 1$:
\begin{align}
(|x_j|/r_\mathrm{F}) \Gamma_{j}&= - \sqrt{\frac{\pi i}{2}} \varphi_0 
 \quad\mathrm{for\ }\varphi_0 \ll 1,\ n=1
 , \\
 \label{eq:NormalizedBrightnessFutureFiducialModel}
(|x_j|/r_\mathrm{F})^2 \left| \Gamma_{j} \right|^2 &=  \frac{\pi}{2} \varphi_0^2
 . 
\end{align}
The approximation is about 14\% high for $n=1$, $\varphi_0=1$, but better for smaller values of $\varphi_0$.
The maximum of $(|x_j|/r_\mathrm{F}) |\Gamma_{j}|$ lies at $n=0.837...$,
where $\tan(n\pi)=\frac{\pi}{2n} n^2-1$, but the maximum value is nearly the same as at $n=1$.

\subsection{Superposition of Neighboring Strips}\label{sec:NeighboringStrips}

An observer will find that 
strips at different offsets from the line of sight produce features 
at distinct positions in the secondary spectrum only if the observations possess sufficient resolution in delay or rate.
If not, the strips will lie within a single observational resolution element, and the resulting points and arclets will be superposed in the secondary spectrum.
The points on the primary arc will lie within a single pixel of the secondary spectrum.
They will contribute to the field with different phases, constructively or destructively, but cannot be distinguished.
We refer to the observed structure from such superposition as a ``feature'' of the secondary spectrum.

\subsubsection{Observational Resolution in the Screen Plane}\label{sec:ObservationalResolution}

For most present observations, 
the limiting observational resolution is insufficient to distinguish strips that lie within about 100 Fresnel zones of one another on the screen.
As discussed in Section\ 3.2.3 of  \citetalias{Gwinn2019},
a structure of width $W$ on the screen (such as a pair of nearby strips),
at a mean offset $\bar x$,
will be unresolved in the secondary spectrum only if:
\begin{alignat}{3}
\frac{B}{\nu_0} &<\frac{4\pi r_\mathrm{F}^2}{ \left| \bar x\right| W}  &&=\frac{2}{n_\mathrm{W}}
,
\\
\mathrm{ and\ } \nonumber \\
\frac{V_x T}{\left| \bar x\right|} &<\frac{4\pi r_\mathrm{F}^2}{ \left| \bar x\right| W}&&= \frac{2}{n_\mathrm{W}}
.
\end{alignat}
where $n_\mathrm{W}=(|\bar x |/2\pi r_\mathrm{F}^2) W$ is the number of Fresnel zones spanned by the width $W$.
Narrower structures appear within a single pixel of the secondary spectrum.
We thus find the number of Fresnel zones spanned by a single pixel:
\begin{align}
\label{eq:rntaunf}
n_\tau &= \frac{2 \nu_0}{B} \\
n_f &= \frac{2  \left| \bar x\right|}{V_x T}
\end{align}
where $n_\tau$ is the number spanned by one resolution element in the delay dimension,
and $n_f$ is the number spanned by one resolution element in the rate direction.
Whichever of these is smaller sets the limiting observational resolution, in pairs of Fresnel zones:
\begin{align}
n_\mathrm{obs} &= \min \left\{ n_\tau, n_f \right\}
.
\end{align}
This number of pairs of Fresnel zones in a corresponds to the linear resolution in the screen plane provided by the observations:
\begin{align}
\label{eq:LinearObservationalResolution}
W_\mathrm{obs} &= \frac{2\pi r_\mathrm{F}^2 }{ \left| \bar x\right|} n_\mathrm{obs}
\end{align}
Thus, if two strips at $x_k$ and $x_\ell$ lie within separation $|x_k - x_\ell |< W_\mathrm{obs}$,
they will combine to form a single feature. We take $ \bar x = \frac{1}{2}\left( x_k + x_\ell\right)$ as the mean offset.

For the observations of pulsar B0834+06 gathered at Arecibo Observatory by \citet{2011ApJ...733...52G}, discussed in Section\ \ref{sec:Observations} below, 
the observing frequency was $\nu_0 =322.495\ \mathrm{MHz}$ and the bandwidth was $B=6.25\ \mathrm{MHz}$.
The observation spanned 4430\ s, with spectra gathered every\ 10\ s.
The Fresnel scale was $r_\mathrm{F} = 8.1\times 10^{10}\ \mathrm{cm}$.
Thus, the observational resolution in delay and rate is:
\begin{align}
\label{eq:GwinnValuesForntaunf}
n_\tau &= 103 \\
n_f &= 3.4 \frac{ \left| \bar x\right|}{r_\mathrm{F}}
\nonumber
\end{align}
The observed primary arc extended from $ | \bar x| < r_\mathrm{F}$ out to $ | \bar x|=800 r_\mathrm{F}$, with the upper limit set by the resolution of the spectrometer.
For $ | \bar x|>31\ r_\mathrm{F}$, the delay resolution $n_\tau$ is the more restrictive limit and sets the observational resolution.
For $ | \bar x|<31\ r_\mathrm{F}$, a quite small segment near the apex, the rate resolution $n_f$ is more restrictive.
Thus, observed features in the secondary spectrum can easily be the superposition of many strips.

For the observations of pulsar B0834+06 on the Arecibo-Green Bank baseline by \citet{2010ApJ...708..232B}, 
$B \approx  0.025 \nu_0$, $|V_x | T \approx r_{\mathrm{F}0}$, and $ | \bar x|/r_\mathrm{F}$ is as great as 2000 and as small as 100 for distinguishable features, although a continuum of features extends to $ | \bar x|/r_\mathrm{F} \approx 1$. 
Thus, $n_\tau=80$ and $n_f = 2  | \bar x|/r_\mathrm{F}$. 
Again, delay resolution determines the observational resolution for all but the innermost portion of the arc.

\subsubsection{Superposition of strips within a resolution element}\label{sec:SuperpositionOfStrips}

If a group of $N_\mathrm{g}$ narrow strips lies within a single observational resolution element, the strips will combine at the observer to form single observed points on the primary arc and secondary arclet. The group will contribute as a single structure. 
If the strips are narrow enough to contribute to the field, and if they are distributed randomly across multiple Fresnel zones, their geometric phases are random and they will combine incoherently.
The resultant intensity will increase proportionately to $\sqrt{N_\mathrm{g}}$, and the net brightness proportionately to ${N_\mathrm{g}}$,
as Equation\ \ref{eq:AnalyticIntensity} indicates. If their locations are periodic with the Fresnel period, they combine coherently, and net brightness is proportional to $N_\mathrm{g}^2$.
On the one hand, this seems unlikely to happen by chance;
on the other, if plasma effects produce periodic structures naturally, the Fresnel period will act as a powerful filter to select them at values of $x_j$ where Fresnel and plasma periodicities match.

We consider a simple model for a group of strips within an observational resolution element.
We suppose that the group contains strips that follow our cosine model for screen phase, described in Section\ \ref{sec:CosineStripModelIntro}.
All of the strips have the same width $w$.
The distribution of the amplitude of screen phase, $\left\{\varphi_{0k}\right\}$, contains negative and positive $\varphi_0$ (excessive and deficient electron column), with zero mean.
The screen phase from the group is:
\begin{align}
\label{eq:VarphiNet}
\varphi_\mathrm{net}(\theta) &= - \sum_{k=1}^{N_\mathrm{g}} \varphi_{0k} \cos(\theta-\theta_k)
,
\end{align}
where, as in\ \ref{eq:ThetaDef}, $\theta$ plays the role of offset $x$:
\begin{alignat}{2}
\theta &= \frac{2\pi}{w} u = \frac{2\pi}{w} (x-\bar x), \quad\quad & \theta_k &= \frac{2\pi}{w} (x_k - \bar x)
,
\end{alignat}
and $\bar x$ lies at the center of the resolution element.
We suppose that the group of strips extends to fill to the observational resolution:
\begin{align}
\Delta\theta &= \mathrm{max}\left\{ \theta_k\right\} -  \mathrm{min}\left\{ \theta_k\right\} = 2 \pi n_\mathrm{obs}
,
\end{align}
as is expected if the distribution of strips in the screen plane is locally uniform (see\ Section\ \ref{sec:Intermittency} below).

We suppose that all of the strips have small screen phase $|\varphi_{0k}| < 1$.
So long as the net screen phase\ \ref{eq:VarphiNet} is also small, we can treat the problem as a linear superposition of the individual strips.
We find the generalization of \ref{eq:OffsetSlopePhi0SmallAnalyticFixXj} to a group of strips:
\begin{align}
\label{eq:OffsetSlopePhi0SmallAnalyticFixXjSum}
(\bar x/r_\mathrm{F}) \Gamma_\mathrm{net}  &=  \sqrt{\frac{2 i}{\pi}} \frac{\sin(n \pi)}{n^2-1}  \sum_{k=1}^{N_\mathrm{g}}  \varphi_{0k} e^{i n \theta_k} 
\ \  \mathrm{for\ }n\ne 1,\ \varphi_0< 1,
\end{align}
where the normalized width is $n=( \bar x/2\pi r_\mathrm{F}^2)\, w$.
We ignore curvature of the Fresnel phase over the group.
Clearly, \ref{eq:OffsetSlopePhi0SmallAnalyticFixXjSum} takes the form of an envelope, equal to the contribution of a single strip as given by \ref{eq:OffsetSlopePhi0SmallAnalyticFixXj}, times a modulation factor given by the sum over $k$.
The envelope is the intensity of a single strip. In particular, it contributes nothing if the strips are too wide, $n>2$.

For a large number of strips, distributed randomly, the modulation factor in \ref{eq:OffsetSlopePhi0SmallAnalyticFixXjSum} has statistics of a 
random walk in the complex plane, with mean of zero and variance of $N_\mathrm{g} \langle \varphi_{0k}^2 \rangle$.
Here, the angular brackets $\langle ... \rangle$ denote an average over the ensemble of statistically-identical realizations of the phase amplitudes $\{ \varphi_{0k} \}$ and offsets $\{ \theta_k \}$.
The distribution of brightness $| \Gamma_\mathrm{net}|^2$ is that of the square modulus of two-dimensional random walk. It is exponential, with maximum probability at $|\Gamma_\mathrm{net} |^2=0$.
The mean of the exponential distribution is $N_\mathrm{g}$ times the brightness of a single strip:
\begin{align}
\label{eq:MeanStrengthOfAGroupSmallFi0}
\left\langle (\bar x/r_\mathrm{F})  \left| \Gamma_\mathrm{net}\right|^2 \right\rangle &= \frac{2}{\pi} \frac{\sin^2(n\pi)}{(n^2-1)^2} N_\mathrm{g}  \langle \varphi_{0k}^2 \rangle
\quad \mathrm{for\ }n\ne 1,\ \varphi_0< 1,
\end{align}
where again $n$ plays the role of a scaled $w$.
In the important special case $n=1$, we must use\ \ref{eq:NormalizedBrightnessFutureFiducialModel}:
\begin{align}
\label{eq:NormalizedBrightnessFiducialStripNeq1SmallFi0}
\left\langle (\bar x/r_\mathrm{F})  \left| \Gamma_\mathrm{net}\right|^2 \right\rangle &= \frac{\pi}{2} N_\mathrm{g}  \langle \varphi_{0k}^2 \rangle
\quad \mathrm{for\ }n = 1{\ and\ }\ \varphi_0< 1.
\end{align}
The mean brightness characterizes the distribution:
13.5\% of the realizations will have more than twice the mean brightness, and 5.0\% more than three times, but 39.3\% less than half.
The median brightness is 69.3\% of the mean.

The exponential distribution of brightness, characterized by \ref{eq:MeanStrengthOfAGroupSmallFi0} and\ \ref{eq:NormalizedBrightnessFiducialStripNeq1SmallFi0}, remain accurate as long as the net screen phase remains less than 1; this is roughly equivalent to $N_\mathrm{g} \langle \varphi_{0k}^2 \rangle / n_\mathrm{obs} <1$. 
If the net screen phase exceeds 1, the linear superposition assumed by \ref{eq:OffsetSlopePhi0SmallAnalyticFixXjSum}-\ref{eq:NormalizedBrightnessFiducialStripNeq1SmallFi0} no longer holds.
In that nonlinear regime, strips of width $w$ are almost always less efficient in their contribution to $\Gamma_\mathrm{net}$ than they are in the linear regime, 
as comparison of\ \ref{eq:AnalyticalModelExactBesselFunctionN}, \ref{eq:AnalyticalModelExactBesselFunction0},
and\ \ref{eq:ResultPhi0SmallAnalyticForm} suggests. The brightness of the feature in the secondary spectrum will be less than \ref{eq:MeanStrengthOfAGroupSmallFi0} would predict.          

\begin{figure}
\centering
\includegraphics[width=0.47\textwidth]{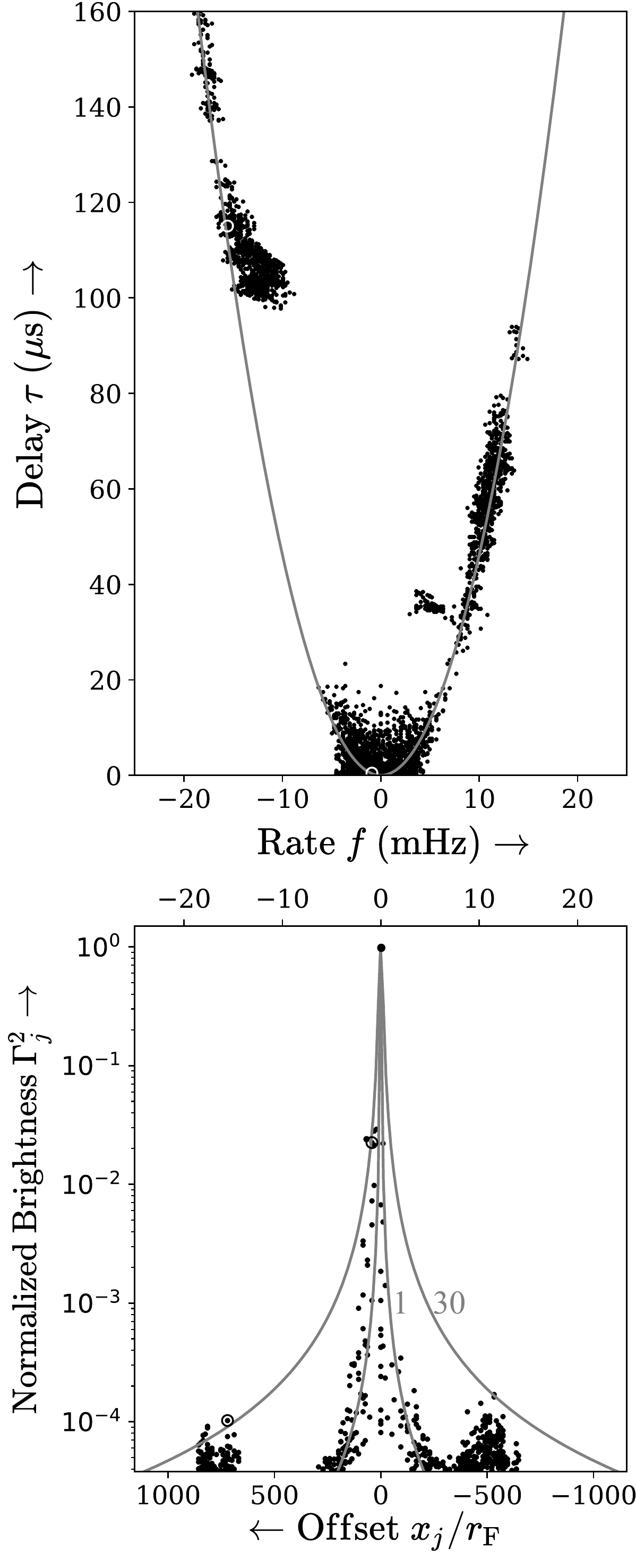} 
\caption{Scintillation arc and brightness. Upper panel: Secondary spectrum from\ \citet{2011ApJ...733...52G}, showing all points with $|\Gamma_j|^2 \ge 4\times 10^{-5}$.
Gray curve shows the location of the primary arc. Lower panel: Normalized brightness $|\Gamma_j^2|$ for the brightest point at each delay $\tau$,
within 40 pixels in delay of the arc,
for $f>0$ and $f<0$,
plotted with rate and inferred offset $x_j$.
Gray curves indicate the mean $|\Gamma_j^2|$ for fiducial models with $\varphi_0=1$, $w=W_{2\pi}$, and $N_\mathrm{g}=1$ or 30, as indicated. 
Black circles in lower panel and white in upper show the points in the first 2 rows of Table\ \ref{tab:ArcletParameters}.
\label{fig:DistributionFigure}}
\end{figure}

\section{Comparison With Observations}\label{sec:Observations}

\subsection{Introduction: Observations With Theory}\label{sec:IntroObs}

We wish to compare the results of our simple model with observations.
In a secondary spectrum that shows a scintillation arc, one can identify the primary arc, bright points along it, and sometimes individual arclets.
The coordinates of delay $\tau_j$ and rate $f_j$, and brightness $|I_j|^2$, characterize each point along the primary arc.
To analyze observations, we determine $x_j$ 
from $\tau_j$ or $f_j$, with\ \ref{eq:DefineAlphaBetaGamma} and parameters $D$, $M$, and $\nu_0$.
We estimate the normalized brightness as $|\Gamma_j|^2 =|I_j|^2/ I_\mathrm{NS}^2$.
The squared mean intensity of the source provides a good estimate of $I_\mathrm{NS}^2$.
The squared mean intensity of the source appears at the origin of the secondary spectrum,
and contains most of the power in the secondary spectrum \citep[see][ and Section\ \ref{sec:BrightnessDistribution} below]{2011ApJ...733...52G}.
From $x_j$ and $|\Gamma_j|^2$ we infer possible combinations of strip phase $\varphi_0$, width $w$, and number of strips $N_\mathrm{g}$ within a resolution element that might contribute to each point on the primary arc, and in principle to each point along the secondary arclets.

A number of instrumental factors complicate extraction of model parameters from the data.
The value at the origin, used to normalize the brightness, may be biased by quantization during digitization
\citep{1998PASP..110.1467J,2004PASP..116...84G,2006PASP..118..461G,2013ApJ...765..135J}.
Pixels close to the rate axis may be corrupted by effects that vary over the observing bandwidth but not with time, such as the shape of the passband. 
Pixels close to the delay axis may be corrupted by effects that vary with time but not over the observing bandwidth, such as
pulse-to-pulse variations in the pulsar's flux density.

\begin{table*}
\centering
 \caption{Inferred parameters for selected points on the primary scintillation arc for pulsar B0834+06, as observed by \citet{2010ApJ...708..232B} and \citet{2011ApJ...733...52G}.}
 \label{tab:ArcletParameters}
 \begin{tabular}{lccccccc}
  \hline
                   &                  &                                    & Normalized                 & Width of a Pair      & Normalized                          & & Mean Fiducial \\
Parameter: & Delay        &Offset                          & Offset                          & of Fresnel Zones  & Brightness                            & & Strips in Group \\
Symbol:     & $\tau_j$     & $x_j$                          &  $|x_j|/r_\mathrm{F}$  &  $W_{2\pi}$           & $\left| \Gamma_j\right|^2$    & $(x_j/r_\mathrm{F})^2 \left|\Gamma_j\right|^2$ & $N_\mathrm{X}$ \\
Units:         & $\mu$s      & cm                              &                                     &  cm                       &                                              &  &   \\
\hline
\citet{2011ApJ...733...52G}: \span\omit\span\omit \span\omit \\
                  & 0.48           & $ 3.4\times 10^{12} $ & 42.                                                   & $1.2\times 10^{10}$ & $ 2.2\times 10^{-2} $ &39. &  25. \\
                  & 115.           & $ 5.8\times 10^{13} $ & 720.                                                 &  $7.1\times 10^{8}$  & $ 1.0\times 10^{-2} $ & 53. & 34. \\
\citet{2010ApJ...708..232B}: \span\omit\span\omit \span\omit \\
 		&  930.          & $1.6\times 10^{14}$    & $2.0\times 10^3$                             & $2.6\times 10^8$      &  $4.4\times 10^{-7}$    & 1.7 & 1.1  \\
                 & 1000.         & $-1.7\times 10^{14}$   & $-2.1\times 10^3$                           & $2.7\times 10^8$      &  $3.9 \times 10^{-6}$    & 16. & 10. \\
   \hline
 \end{tabular}
\end{table*}

\subsection{Observed Brightness with Offset}\label{sec:BrightnessDistribution}

\subsubsection{Formation of the figure}\label{sec:FormationFigure}

The upper panel of Figure\ \ref{fig:DistributionFigure} shows a fairly typical secondary spectrum, for pulsar B0834+06, gathered at Arecibo Observatory, from \citet{2011ApJ...733...52G};
full details of the observation are presented there.
The upper panel shows all pixels above an empirical noise cutoff of $|\Gamma_j|^2 \ge  4\times 10^{-5}$.
The region along the rate axis was eliminated, except within 40 pixels in rate of the origin, to exclude instrumental effects as described above.
The curve shows the estimated position of the primary arc;
its location is indistinguishable from the results of \citet{2010ApJ...708..232B} for the same pulsar at nearly the same observing frequency, but at a different epoch and with different telescopes.

The lower panel of Figure\ \ref{fig:DistributionFigure} shows points selected from the upper panel to show the brightness of the primary arc.
For these points, we selected the brightest point at each delay, at both positive and negative rate.
To reduce effects of off-arc features, we considered only points within 40 pixels of the primary arc in delay ($6.4\ \mu\mathrm{s}$).
Our selection may not include all points on the primary arc: for example, some points on a secondary arclet may be brighter than an unrelated primary point.
However, it includes the brightest primary points, and provides at least an upper limit on the brightness of the rest.
We projected the selected points onto the rate axis and plotted their normalized brightness $| \Gamma_j|^2$ with rate and offset on the lower panel. 
We calibrated the rate $f_j$ as offset $x_j = f_j/\beta$ using \ref{eq:DefineAlphaBetaGamma} and the parameters given in \citetalias{Gwinn2019}.
We display both rate and offset as dual horizontal axes on the plot;
because $\beta<0$, the two horizontal axes increase in opposite directions.

As Figure\ \ref{fig:DistributionFigure} shows, 
the pixel at the origin of the primary spectrum, at the peak of the lower plot, is more than 30 times brighter than the rest, as expected for an undeflected line of sight subject to weak or no scattering.
Because of the limited frequency resolution of the spectrometer,
the maximum observable delay is $\tau = 160\ \mu \mathrm{s}$.
The arc appears to extend well beyond this limit.
\citeauthor{2010ApJ...708..232B} observed arcs with delays as great as 1 ms for this pulsar, using a much higher-resolution software correlator.
At this observing epoch, the arc shows a strong ``core'' of points at small offset, and weaker ``clumps'' further out on the arms of the arc.
``Gaps'' empty of bright features appear between the clumps and the core.
Off the primary arc, a small arc appears at $\tau=35\ \mu\mathrm{s}, f=5\ \mathrm{mHz}$, and a series of arcs near $\tau=105\ \mu\mathrm{s}, f=-10\ \mathrm{mHz}$. ``Canted'' noodles can produce these, as discussed in \citetalias{Gwinn2019}.
Other studies commonly show similar clumping along the primary arc, and occasional off-arc structures.

Table\ \ref{tab:ArcletParameters} gives detailed parameters for two of the points shown in Figure\ \ref{fig:DistributionFigure}.
These points are marked with circles: black in the lower figure, white in the upper. 
To measure the parameters in the table, we located the brightest pixel in the neighborhood,
and adopted its brightness, delay and rate to calculate $x_j$ and $|\Gamma_j|^2$.

The table also includes two features at $\tau \approx 1000\ \mu\mathrm{s}$, from the remarkable work of \mbox{\citet{2010ApJ...708..232B}}.
This observation attained extremely high delays by use of a software correlator with an extremely large number of channels.
The 1-ms arclet is remarkably strong for its separation from the origin, but has nontrivial structure. It does not lie on the primary arc; 
\mbox{\citet{2016MNRAS.458.1289L}} and \mbox{\citet{2018MNRAS.478..983S}} argue that two screens at different distances are involved in its formation.
Thus, it may be anomalous.
The arclet at delay $\tau = 930\ \mu\mathrm{s}$ is an order of magnitude weaker but is more typical, and has a simpler structure.
These features were relatively bright, but were observed at high delay.
Some weak arclets may have been detected at up to $\tau = 1200\ \mu\mathrm{s}$, but not much further than that.
No instrumental factors prevented observations of features at much higher delays, out to almost 2\ ms; but none were detected.
Thus, $\tau \approx 1\ \mathrm{ms}$ may represent a physical limit to the delay, set by the scattering process.

\subsection{Models}

\subsubsection{Fiducial strip}\label{sec:FiducialStrip}

Figure\ \ref{fig:DistributionFigure} shows the brightness for two forms of our cosine model, as two curves.
The inner curve, marked ``1,'' shows the brightness for a single strip within a resolution element, with screen phase $| \varphi_0 | = 1\ \mathrm{radian}$,
and width equal one pair of Fresnel zones: $w=W_{2\pi}$. 
As discussed in Sections\ \ref{sec:CosineStripModel} and\ \ref{sec:BrightnessInTimeWidthSuperposition}, 
such strips are efficient scatterers for their phase amplitude and are approximately linear in superposition. 
Such a strip produces brightness given by\ \ref{eq:NormalizedBrightnessFiducialStripNeq1SmallFi0} with $\varphi_0=1$:
\begin{align}
\label{eq:FiducialStripBrightness}
 \Gamma_\mathrm{fid}^2 \equiv \left(\frac{\pi}{2}\right) \left(\frac{r_\mathrm{F}}{| x_j|}\right)^2
.
\end{align}
We call this a ``fiducial strip.''
The curve ascends as $\Gamma_\mathrm{fid}^2 \propto |x_j|^{-2}$ because $W_{2\pi} \propto |x_j|^{-1}$.

The outer curve, marked ``30,'' shows a brightness of $30\, \Gamma_\mathrm{fid}^2$.
A group of $N_\mathrm{g}=30$ fiducial strips within each resolution element with $\langle \varphi_{0k}^2\rangle=1$, combining incoherently, will produce this mean brightness.
The observational resolution spans 103\ pairs of Fresnel zones, for $|x_j|/r_\mathrm{F}>31$ (\ref{eq:GwinnValuesForntaunf});
thus, any resolution element could easily hold 30 fiducial strips.
In the following sections and Table\ \ref{tab:ArcletParameters}, we describe the brightness ${|\Gamma_j|^2}$ of a point $j$ on the primary arc
in terms of the brightness of one fiducial strip,
using the parameter:
\begin{align}
\label{eq:FiducialStripMultiplicity}
N_\mathrm{X} \equiv \frac{|\Gamma_j|^2}{\Gamma_\mathrm{fid}^2} =\left(\frac{2}{\pi}\right)(x_j/r_\mathrm{F})^2 \left|\Gamma_j\right|^2
.
\end{align}
Note that $N_\mathrm{X}$ is related to the effective area for the feature: $A_j = N_\mathrm{X} A_\mathrm{fid}$,
where $A_\mathrm{fid} = 2\pi r_\mathrm{F}^2 \Gamma_\mathrm{fid} = \pi^2 r_\mathrm{F}^4/|x_j|^2$.

\subsubsection{Three models}\label{sec:3Models}

One can imagine at least 3 classes of noodle models for the points on the primary arc described by Figure\ \ref{fig:DistributionFigure} and Table\ \ref{tab:ArcletParameters}.
We will classify these by the number of strips within a resolution element $N_\mathrm{g}$, as compared with the brightness parameter $N_\mathrm{X}$ given by\ \ref{eq:FiducialStripMultiplicity}.

In the ``single-strip model,'' $N_\mathrm{g} = 1$, so we assume that each point on the primary spectrum is associated with a single strip on the screen.
Thus, each observational resolution element contains only one strip.
The width $w$ and phase amplitude $\varphi_0$ of that strip depend on the brightness and location of the strip.

In the ``group'' model, $N_\mathrm{g} = N_\mathrm{X}$ strips lie in the resolution element. They have widths of one pair of Fresnel zones, $w=W_{2\pi}$, but have a distribution of phase amplitudes,
with mean $\langle \varphi_0\rangle=0$ and variance $\langle \varphi_0^2\rangle=1$. The strips sum incoherently to produce mean brightness equal to the observed brightness, as Section\ \ref{sec:SuperpositionOfStrips} describes.
For points with $N_\mathrm{X}<1$, we invoke a single strip with screen phase $| \varphi_0| <1$.

In the ``population'' model, a population of $N_\mathrm{g} \gg N_\mathrm{X}$ strips within the resolution element sums incoherently to produce the observed brightness.
The strips have widths $w=W_{2\pi}$.
The strips again have a distribution of phase amplitudes,
with mean of zero and variance of $\langle \varphi_0^2 \rangle \approx N_\mathrm{X} / N_\mathrm{g} $.
If the feature is particularly bright relative to its neighbors, then we may suppose that it is on the high end of the exponential distribution of brightness described in Section\ \ref{sec:SuperpositionOfStrips}:
it is ``lucky''.
For example, if it is brighter than $1/5\%=20$ neighbors, it may by chance lie at 3 times the mean, so that the model requires only $1/3$ as many strips or $1/3$ as large variance of screen phase $\left\langle \varphi_0^2 \right\rangle$ as\ \ref{eq:NormalizedBrightnessFiducialStripNeq1SmallFi0} would predict, to produce the observed brightness.
This model requires only strips of low phase amplitude, at the cost of large numbers of strips within a resolution element.

\subsubsection{Single-strip model}

\begin{figure}
\centering
\includegraphics[width=0.45\textwidth]{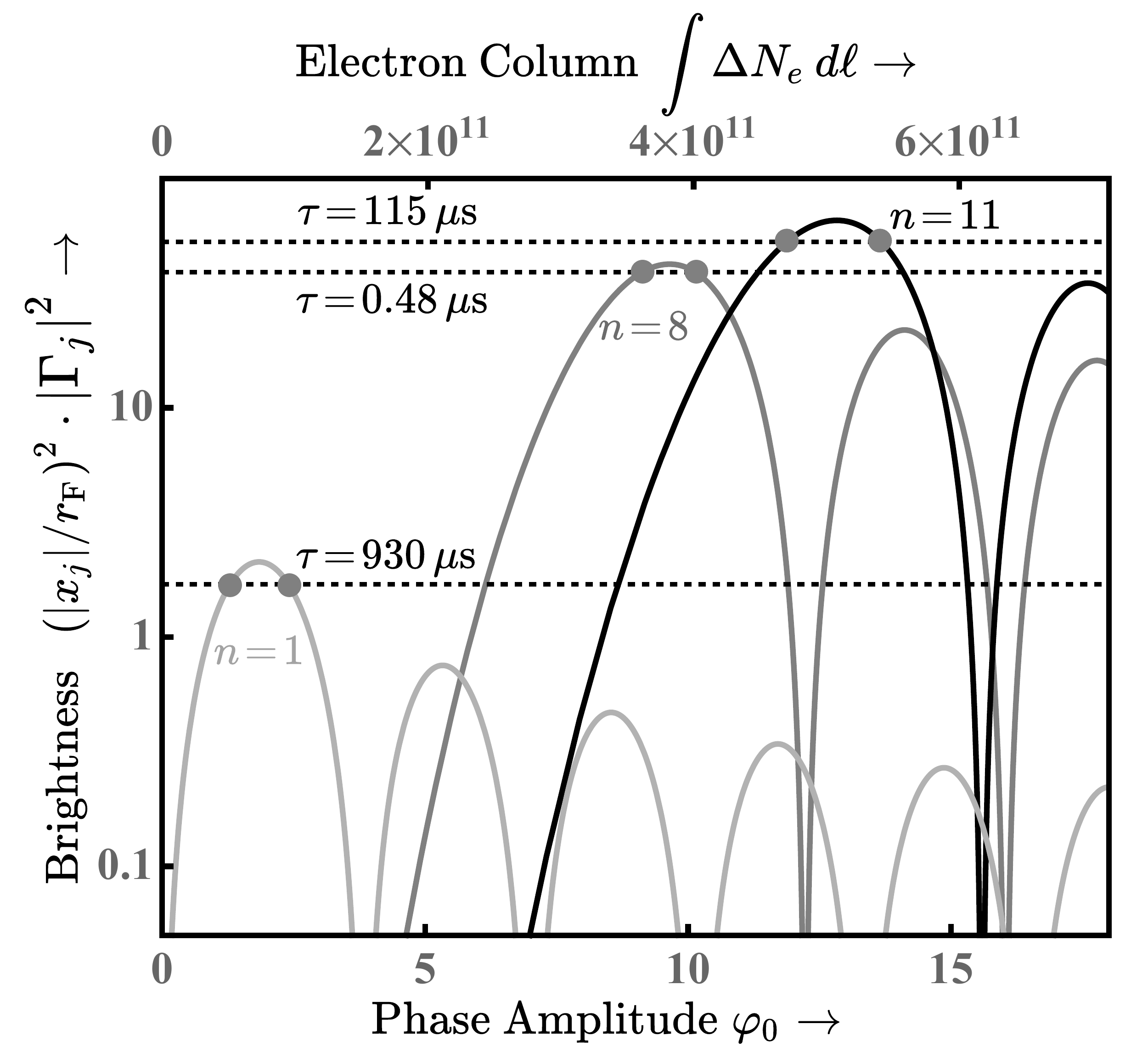} 
\caption{
Single-strip model solutions for strip phase $\varphi_0$ for features at $\tau=0.48\ \mu\mathrm{s}$, $\tau=115\ \mu\mathrm{s}$, and $\tau=930\ \mu\mathrm{s}$.
Upper horizontal axis shows electron column corresponding to $2\varphi_0$.
\label{fig:AnalyticalModelColumns}}
\end{figure}

A single-strip model 
requires inversion of \ref{eq:Gammaj_from_SnjSnj0_xjForm}, using the observed values for $| \Gamma_j |^2$ and $x_j$, while allowing $\varphi_0$ and $n$ to vary; 
The observed $| \Gamma_j |^2$ and $x_j/r_\mathrm{F}$ define a plane perpendicular to the vertical axis in Figure\ \ref{fig:SameViewPanel2_Vary_w}.
Any point along the intersection of that plane and the surface in the figure is a possible solution.
In this case, $n = (|x_j|/2\pi r_\mathrm{F}^2)\,w$ corresponds to a normalized width.
For simplicity, here we specialize to integral values of $n$, so that we explore the solid-line `skeleton'' in Figure\ \ref{fig:SameViewPanel2_Vary_w}. 
Figure\ \ref{fig:AnalyticalModelColumns} displays vertical sections through Figure\ \ref{fig:SameViewPanel2_Vary_w}, at selected integer values of $n$, on a semilog scale.
The $\tau=115\ \mu\mathrm{s}$ feature requires $n\ge 10$.
The figure displays the curve for $n=11$,
with solutions at $\varphi_0=11.88$ and $13.67$.
The $\tau=0.48\ \mu\mathrm{s}$ feature requires $n\ge 8$. Even though it is brighter than the $\tau=115\ \mu\mathrm{s}$ feature, it  can be fit with smaller width in Fresnel zones (although not in cm), and with lower screen phase $\varphi_0= 9.12$ or $10.14$. This is because it lies at smaller offset, where Fresnel zones are wider.
The $\tau=930\ \mu\mathrm{s}$ feature can be fit with $n=1$ and $\varphi_0 = 1.29$ or $2.40$.
Wider strips will also fit in each case, although only with larger $\varphi_0$.
For example, as the figure shows, the $\tau=0.48\ \mu\mathrm{s}$ and $\tau=930\ \mu\mathrm{s}$ features can also be fit with $n=11$, but only with $\varphi_0>8$.
Even larger values of $\varphi_0$ and $n$ also provide solutions, as Figure\ \ref{fig:SameViewPanel2_Vary_w} shows.

\subsubsection{Group model}

A fit to the group model for the $\tau=115\ \mu\mathrm{s}$ feature leads to $N_\mathrm{g}=N_\mathrm{X}=34$ incoherently superposed strips, as Table\ \ref{tab:ArcletParameters} states.
In practice, this feature requires the largest number of strips of all those observed, so we might appeal to the randomness of the distribution of brightness from superposition discussed in Section\ \ref{sec:SuperpositionOfStrips}, and assume that this feature is among the top 5\% of chance brightnesses, so that we need only $N_\mathrm{X}/3=11$ strips.
For the  $\tau=0.48\ \mu\mathrm{s}$ feature we require $N_\mathrm{X}=25$ strips, or fewer if we again appeal to chance.
The group model for the $\tau=930\ \mu\mathrm{s}$ feature is the same as for the $N_\mathrm{g} =1$ model.

\subsubsection{Population model}

A fit to the population model with $N_\mathrm{g} \gg N_\mathrm{X}$ can use strips with electron columns smaller than the fiducial value, but requires many more of them.
So, for example, for the $\tau=115\ \mu\mathrm{s}$ feature, we could have $N_\mathrm{g}=3400$ strips with $\langle\varphi_0^2 \rangle = 0.01$.
Again, assumption of a more-coherent superposition with 5\% chance would allow us to use only $1100$ such strips. 
Similar considerations hold for the $\tau=930\ \mu\mathrm{s}$ feature; we could, for example, create a model with $N_g=11$ strips with $\langle\varphi_0^2 \rangle = 0.01$,
or a ``lucky'' model with $N_g=800$ strips and $\langle\varphi_0^2 \rangle = 10^{-4}$. Of course, an infinite number of combinations of $N_\mathrm{g}$,  $\langle\varphi_0^2 \rangle$, and even $w$ are possible.

\section{Discussion}\label{sec:Discussion}

Additional constraints or additional observations can illuminate the three models of Section\ \ref{sec:3Models}.
Among the possible constraints are minimal electron column, or electron density;
a match to the observed brightness with offset that arises naturally from features of the model;
and a match to theoretical expectations for the structure of a turbulent plasma.
As we discuss in this section, the population model seems to fit these constraints most elegantly.
Further observations of the variations of brightness of features with time and observing frequency, and the brightness distribution they trace out, can qualify or disqualify models.

\subsection{Electron density}\label{sec:ElectronDenstiyVariationsFromNoodles}

\subsubsection{Single-noodle model}

In the single-noodle model, phase amplitude provides the electron column for a strip $\int \Delta N_e\cdot d\ell $.  
The fluctuation of electron density $\Delta N_e$ in a physical structure, a ``noodle,'' depends on the extent of the noodle along the line of sight.
The maximum phase in a strip, in the cosine model is $2\varphi_0$.
From \ref{eq:basic_plasma_phase}, we find the peak electron column for a fiducial strip:
\begin{align}
\left[ \int \Delta N_\mathrm{e}\cdot d\ell \right]_\mathrm{fid} &=7.8 \times 10^{10}\ \mathrm{cm}^{-2}  
\end{align}
For approximately cylindrical filaments (``spaghetti'') at an angle $\eta$ to the screen, 
the typical electron density is \citepalias[][Section\ 5.1]{Gwinn2019}:
\begin{align}
\Delta N_\mathrm{e} &\approx \frac{\int \Delta N_e\cdot d\ell}{w \csc \eta} 
.
\end{align}
For sheets (``lasagna''), the extent along the line of sight depends on how far the sheets follow the line of sight.
\citet{2018MNRAS.478..983S} define a ``radius of curvature'' $R$ that effectively sets this distance.
For our purposes here, $w\csc \eta$ and $R$ are equivalent.
As \citet{2014MNRAS.442.3338P} point out, observations strongly select for $\eta\ll 1$, or $R\gg r_\mathrm{F}$.

\subsubsection{Group and population models}

The group model and the population model lower electron density $\Delta N_\mathrm{e}$ further through superposition of features within a resolution element.
The group model lowers it by the fixed factor $1/\sqrt{N_\mathrm{X}}$, and the population model by an arbitrary factor of $1/\sqrt{N_\mathrm{g}}$.
Both models can lower the electron column by diluting it over a resolution element, by a factor of about $N_\mathrm{X}/n_\mathrm{obs}$.

In the population model, the required fluctuation of electron density can be well below the local average density.
For example, a factor of $\csc \eta = w/R = 100$ and $N_\mathrm{g}=10^6$ lead to $\Delta N_\mathrm{e} \approx 0.003\ \mathrm{cm}^{-3}$
for the $930\ \mu\mathrm{s}$ feature, less than the $N_\mathrm{e}\approx 0.25\ \mathrm{cm}^{-3}$ expected for the scattering regions \citep{1990ApJ...353L..29S}. 
If a small fluctuation of electron column is considered desirable, the population model is a clear favorite.
Optics only weakly constrains the number of superposed noodles [][Section\ 5.1]{Gwinn2019}; indeed, the fractal nature of turbulence would suggest that many noodles are to be expected.

\subsection{Overall distribution of brightness}

A strip of fixed width $w$ and phase amplitude $|\varphi_0|<1$ will contribute only when $w<W_{2\pi}$, as Figure\ \ref{sec:SmallElectronColumn} shows.
This acts as a Fresnel filter that strongly affects the relation between the distribution of strips in parameter space $\{ \varphi_0, w, N_\mathrm{g}\}$ and the distribution of brightness with offset.
A single such strip will trace out the light curve displayed in Figure\ \ref{sec:SmallElectronColumn}.
A distribution of such strips with uniform spacing will produce a more sharply-peaked distribution of brightness that extends over the same range.
This is because the width of a resolution element $W$, in units of length, varies inversely with $|\bar x|$ (\ref{eq:LinearObservationalResolution}),
so that the number of strips within a resolution element varies as $N_\mathrm{g} \propto |\bar x|^{-1}$.
By contrast, the observed distribution of brightness is still more sharply peaked: it has a form closer to $|\Gamma_j|^2 \propto |x_j|^{-2}$, as Figure\ \ref{fig:DistributionFigure} shows.
Thus, to reproduce the observed distribution of brightness requires a model including cohorts of strips with different widths, that are visible out to different maximum offsets.

Theories of turbulence predict power-law forms for the correlation of density with spatial separation in a turbulent fluid \citep{landau2013fluid,frisch1995turbulence,1995ApJ...438..763G,2001ApJ...562..279L}.
More precisely, they predict the second-order structure function of velocity, $\langle |\mathbf{V}(\mathbf{x} + \Delta \mathbf{x}) - \mathbf{V}(\mathbf{x} )|^2\rangle_\mathbf{x}$, as a function of $\Delta\mathbf{x}$.
From this, one can infer the power-law form of the analogous structure function of density.
A power-law form for the structure function of density would suggest a power-law form for the average brightness;
this is roughly compatible with the distribution in Figure\ \ref{fig:DistributionFigure}.
The observations of \citet{2010ApJ...708..232B} suggest a similar power-law form.
However, the connection between the distribution of brightness and the structure function of electron density is necessarily quite indirect, because the Fresnel filter is so strong, as noted above.
Within noodle models, 
a model distribution of noodles in parameter space $\{ \varphi_0, w, N_\mathrm{g}, x_j \}$ yields the distribution of brightness with position,
as well as the structure function of density with position.
However, the distribution of brightness is clearly not stationary with $x_j$, or with time.
Particularly given the marked changes seen with observing epoch even for individual pulsars, more observations of more pulsars are clearly needed to address the correlation function of density with any degree of accuracy.

\subsection{Clustering: Spatial spectrum and Intermittency}\label{sec:Intermittency}

Features are strongly clustered in the secondary spectrum, indicating clustering in space.
The clumps at larger offsets in Figure\ \ref{fig:DistributionFigure} demonstrate that noodles are clustered on spatial scales larger than the observational resolution. This suggests that they may well be clustered on smaller scales as well, reinforcing the fiducial and population models.
The observed clumps will shift toward positive rate with time, as expected for the noodle model \citep[][Section\ 5.5]{Gwinn2019},
and as documented in observations \citep{2003ApJ...599..457H}. The distribution is not stationary with time.

The clustering of noodles is in accord with ``intermittent'' turbulence, where only a few of the many degrees of freedom of the fluid are active \citep{frisch1995turbulence}.
Simulations, experiments, and observations of strong turbulence show that most of the dissipation takes place in localized, organized structures that fill only a small percentage of the volume \citep{2003ApJ...595..812C,2007ApJ...658..423K,2009ASPC..417..243F,2015RSPTA.37340154M}.
Velocity and density fields are both intermittent, but with markedly different statistics.
The strong clustering of features in the secondary spectrum, and their appearance as organized structures, suggests pronounced intermittency.

Perhaps unfortunately, intermittency further complicates any comparison of brightness and position with the spatial spectrum of turbulence.
The structure function expresses variation of density (or velocity) in terms of second moments.
In principle, second and higher-order structure functions, and moments, can describe intermittency,  
but only very large averages over time as well as space can measure them accurately.
Incomplete averages are strongly biased by rare events.
The situation is similar to describing the outcome of a lottery by moments: the average payoff, mean square payoff, and so on are all dominated by the most uncommon (and largest) prizes.
Changes in the distribution of smaller prizes produce only small differences in all of the moments.
Under the circumstances, a description other than moments may be most useful.
L\'evy statistics, rather than Gaussian random walks, can describe such scattering processes \citep{2003ApJ...584..791B,2003PhRvL..91m1101B,2005ApJ...624..213B}.
Observations at many epochs, with appropriately weighted averaging, may provide insight into intermittent turbulence.

\subsection{Inner scale}\label{sec:InnerScale}
	
The maximum observed delay, at $\tau\approx 930\ \mu\mathrm{s}$, may correspond to the inner scale, the minimum scale of a turbulent cascade.
The physics of turbulent fluids in general, and reconnection regions in particular, involves a great range of spatial scales, ranging from the maximum, outer scale of the entire sheet to a minimum, inner scale.
These small-scale fluctuations may scatter electrons, providing the resistance required for reconnection.
A plausible minimum scale for turbulence is the ion inertial length, where electron and ion fluids decouple \citep{1990ApJ...353L..29S,2019ApJ...878..157X}.
On the other hand, \citet{2001ApJ...562..279L} suggest that the minimum widths of sheets or filaments should be about the ion cyclotron radius. 
For conditions inferred for interstellar scattering plasma, these lengths are both a few hundred km. 
On the basis of observational evidence,
\citet{1990ApJ...353L..29S} propose that the minimum length scale is 50 to 200\ km; \citet{2018ApJ...865..104J} find a value for the minimum scale of turbulence of 800\ km toward the Galactic center source SgrA*. They propose this as the width of one-dimensional structures responsible for turbulence, similar to the noodles considered here.
In the noodle model, the inner scale represents a minimum thickness of the noodles, and a minimum width for strips in the scattering screen.

As noted above, the arclet observed by \citet{2010ApJ...708..232B} at delay $\tau = 930\ \mu\mathrm{s}$ appears typical aside from its enormous delay, and has quite a simple structure.
Another feature at $\tau = 1000\ \mu\mathrm{s}$ has a more complicated structure \citep{2016MNRAS.458.1289L,2018MNRAS.478..983S}, and 
some fainter features may appear near $\tau = 1200\ \mu\mathrm{s}$.
However, 
the observations of \citeauthor{2010ApJ...708..232B}  could have detected features at much greater delays extending out to $\tau = 2000\ \mu\mathrm{s}$, but none were observed.
Thus, $\tau \approx 1000\ \mu\mathrm{s}$ may be a physical limit to the delay, set by the scattering process.
This maximum delay corresponds to a minimum strip width of $w=W_{2\pi} = 2600\ \mathrm{km}$.
The characteristic radius of the strip is probably a better measure than the total width.
Because the fluctuation of electron column is already 0 at the strip edges, and falls by half halfway from its middle to its edge, a reasonable measure of characteristic radius is $w/4 = 650\ \mathrm{km}$.
This is in accord with the other observational and theoretical estimates noted above.
Clearly, further observations at very high delays of pulsar B0834+06 and other pulsars, over a range of observing frequencies, would be most useful in identifying the inner scale.
A particularly interesting case might be pulsar J0437$-$4715, which shows strong scintillation arcs and is among the brightest and closest pulsars \citep{Reardon2018PhDT}. 

\subsection{Comparisons with Other Models}\label{sec:ComparisonWithOtherModels}

On the theoretical side, the noodle model is distinguished from previous models in that it applies wave optics rather than ray optics, and that it may invoke incoherent superposition of multiple structures within an observational resolution element.
These distinguishing features enable the noodle model to produce secondary spectra with much smaller fluctuations of electron column than earlier models. 
The noodle model achieves low electron column by invoking structures that are very narrow compared with previous models: about as wide as a Fresnel zone.

By contrast, in the ray picture of previous models, plasma clouds act as prisms to deflect rays from the source to the observer.
The refracting surface must have effective area $A$ sufficient to produce the secondary spectrum,
analogously to the effective area calculated for the noodle model.
In the ray picture, the effective area is the area of the region over which variation of Fresnel plus screen phase remains small, $\Delta \Phi \lesssim \pi/4$ (see Section\ \ref{sec:StatPhaseApprox}).
Some models invoke approximately spherically-symmetric scatterers, perhaps small clouds of plasma \citep{2004MNRAS.354...43W,2005MNRAS.362.1279W,2008MNRAS.388.1214W}.
A spherical cloud must have diameter of about 70 pairs of Fresnel zones,
to attain effective area that can 
match the observed brightness of the brighter features in Table\ \ref{tab:ArcletParameters}.
Across that region,
the scatterer must provide sufficient phase gradient to cancel the Fresnel phase. Thus, the gradient must accumulate an electron column of more than $2\pi\cdot 70 \approx 440$ radians, or an electron column of about $1.7\times 10^{13}\ \mathrm{cm}^{-2}$. For a spheroidal model, this column must be concentrated in a distance along the line of sight equal to its diameter, or as little as $5\times 10^{10}\ \mathrm{cm}$, for an inferred electron density of $\Delta N_e \approx 350\ \mathrm{cm}^{-2}$. 

\begin{figure}
\centering
\includegraphics[width=0.35\textwidth]{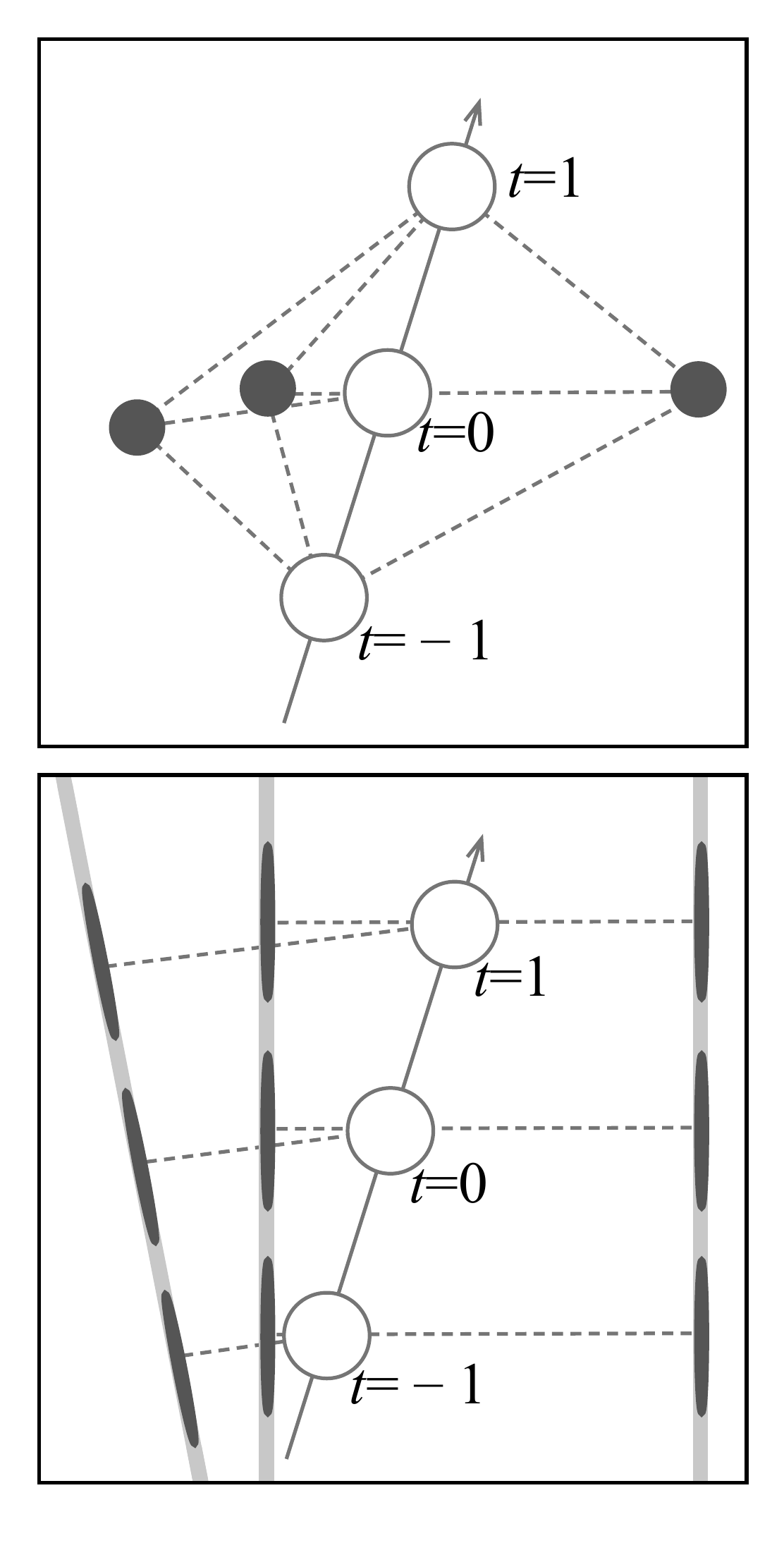} 
\caption{Comparison of spheroid model (upper) with noodle model (lower), showing the locations of the scatterers (dark patches) and the line of sight (open circles). As the source moves with time $t$, the scatterers remain fixed in the spheroid model, but move with the source in the noodle model. Nearly parallel noodles lead to an elongated distribution of scatterers on the sky.
\label{fig:CompareDumplingSpaghetti}}
\end{figure}

Figure\ \ref{fig:CompareDumplingSpaghetti} schematically compares the optics of spheriod and noodle models.
The scattering regions on the different noodles move with the undeflected line of sight;
this provides a natural means of producing the observed elongated distribution of scatterers\ \citep{2010ApJ...708..232B}.
Similarly, because the scattered field of a noodle cancels unless the noodle width is less than two pairs of Fresnel zones (or a few Fresnel zones, for the single-strip model) ensures that the distribution of scatterers is always centered on the image of the source.

Models that invoke corrugated sheet-like scatterers, 
such as those proposed by \citet{2014MNRAS.442.3338P, 2016MNRAS.458.1289L}\ and\ \citet{2018MNRAS.478..983S}, can achieve the required area with a less extensive gradient because optics selects the segment closest to the line of sight, as for the noodle model, and the length of that segment is about $\sqrt{2\pi} r_\mathrm{F}$, as in the $y$-direction in the noodle model.
Sheet models closely resemble the single-strip noodle model of Section\ \ref{sec:3Models}, in the ray-optics limit of Section\ \ref{sec:StatPhaseApprox}. However, sheet models deal with more general profiles of screen phase than the cosine model treated here. Sheet and noodle models occupy nearby, and sometimes overlapping, regions of parameter space.

\subsection{Observational Tests}

Observationally, wide-bandwidth and multi-epoch observations provide an opportunity to test models.
In principle, a direct Fourier transform of $\Gamma_j$ from the frequency domain to the domain of offset in the screen plane $x$ could provide an ``image'' of the screen phase, as discussed in Section\ 2.3.2 of\ \citet{Gwinn2019}.
In effect, from observations at each frequency one obtains the screen phase, modulated by the spatial frequency provided by the Fresnel phase. 
The wide bandwidth and high dynamic range required to make such images may be difficult to obtain in practice,
but single-strip models make simple predictions for frequency behavior.
For group and population models, wideband observations should reproduce the expected exponential distribution of brightness discussed in Section\ \ref{sec:3Models}, if the noodles are numerous and statistics are Gaussian.
If statistics are L\'evy, as discussed in Section\ \ref{sec:Intermittency}, the distribution of brightness may be power-law.

Noodles are narrow enough to evolve visibly within a single observation.
At the fiducial width of one pair of Fresnel zones, the 
crossing time at the Alfv\`en speed of $V_\mathrm{A}\approx 18\ \mathrm{km\ s}^{-1}$ \citep{1990ApJ...353L..29S}
is $W_{2\pi}/V_\mathrm{A} \approx 400\ \mathrm{s}$ for the 
$\tau=115\ \mu\mathrm{s}$ feature in Table\ \ref{tab:ArcletParameters}
and $\approx 150\ \mathrm{s}$ for the 
$\tau=930\ \mu\mathrm{s}$ feature.
On these timescales, one might expect to see transverse Alfv\'en waves pass through the point where the noodle is closest to the undeflected line of sight. 
These waves correspond to lateral displacements of magnetic field lines and plasma. 
This might be visible as a time-varying ``slant'' of the noodle, causing a change in rate, but not in delay, for the resulting point on the primary arc. One might even see more than one such slant, if the wavelength of the Alfv\'en wave was less than the Fresnel scale.

Coherent superposition of strips is effective over relatively narrow bandwidth.
A group of $N_\mathrm{g}$ strips spaced across $N_\mathrm{g}$ adjacent pairs of Fresnel zones will interfere constructively only over bandwidth $B=\nu_0/N_\mathrm{g}$.
Wider-bandwidth observations of individual strips will broaden the resulting features parallel to the rate axis \citep[][Section\ 5.5 and Figure 6]{Gwinn2019},
but groups of coherently-interfering strips will appear strongly only over narrow fractions of that width.
Interestingly, in the extremely wide-bandwidth observations of\ \citet{Reardon2018PhDT}, the scintillation arc of pulsar J0437-4715 do not uniformly fill the expected region of the secondary spectrum;
rather, it has a filamentary structure, perhaps suggestive of incoherent superposition, with the degree of chance coherence varying as a function of observing frequency.

\section{Summary}

In \citetalias{Gwinn2019} , we described how narrow filaments or sheets of plasma, ``noodles,'' 
can act to form the parabolic scintillation arcs seen for many scintillating pulsars.
In this model, scattering takes place in a narrow strip located in a scattering screen,
at an offset $x_j$ from the undeflected line of sight.
The strip is indefinitely long, with electron column varying across the strip but uniform along it.
In this work, we investigate a cosine profile for the deviation of electron column in the strip,
$\varphi(x) = -\varphi_0 (1+\cos(\pi u/w)$, with $u=x-x_j$ the offset from the centerline of the strip at $x_j$, and $w$ the strip width.
Kirchhoff integration yields an analytic expression for the scattered field, when the width of the strip is an integral number of Fresnel zones: $w=n\cdot W_{2\pi}=n\cdot 2 \pi r_\mathrm{F}^2/x_j$,
at the point $x_j$ where the strip is closest to the undeflected line of sight.
The model thus yields the brightness of the resulting feature in the secondary spectrum from the parameters $\{ \varphi_0, x_j, w\}$.
We show that for large electron column, the analytic model matches results of the stationary-phase approximation;
we note that the paths with stationary-phase points are rays, and call that regime ``ray optics''.
For smaller electron column, the approximation is inaccurate, and for the smallest electron columns no stationary-phase points exist,
but the strip scatters radiation efficiently.
We call this regime ``wave optics''.
We calculate the brightnesses of features from strips of different widths, and show the brightest feature for the least fluctuation of electron column appears when the width of a strip is about one pair of Fresnel zones,
at the strip's closest approach to the line of sight: $w=W_{2\pi}$.
For small electron column, so that $|\varphi_0|<1$, such a strip will be visible in the secondary spectrum only when its width is less than about two pairs of Fresnel zones, or at $|x_j| \lesssim 2\cdot 2 \pi r_\mathrm{F}^2/w$;
it will be invisible for greater $x_j$, because the Fresnel phase will cancel across its width.

\citetalias{Gwinn2019} showed that the resolution of the secondary spectrum is finite in the screen plane: for most regions of most present observations, strips within about 100 Fresnel zones cannot be distinguished.
Strips closer than this will combine to form a single point on the primary arc, and single arclet, in the secondary spectrum.
If randomly distributed within a resolution element, and narrow enough to be visible, a group of $N_\mathrm{g}$ strips will combine incoherently to increase the brightness of the resulting feature in the secondary spectrum.
If separated by integral numbers of Fresnel zones, they will combine coherently.
Incoherent superposition results in an exponential distribution of brightness, with mean brightness equal to $N_\mathrm{g}$ times the brightness of a single strip,
and with 5.0\% of the distribution more than three times the mean.

We compare our theoretical model with observations of scintillation arcs by \citet{2010ApJ...708..232B} and \citet{2011ApJ...733...52G}.
We find that 3 types of noodle models can match observations, classified by the number of strips superposed.
Observed features can be matched by a single strip, with $\varphi_0 =12$ radians or less. 
They can also be matched by several strips with screen phase of 1 radian, corresponding to maximum electron column of $7.9\times 10^{10}\ \mathrm{cm}^{-2}$.
Finally, they can be matched by an indefinitely large number of strips with indefinitely small electron column.
These electron columns are much smaller than those found for previous models.
That is because, being extremely narrow in one direction, the noodles do not require a large-scale gradient of screen phase to cancel the Fresnel phase across their width.
Their great lengths in the perpendicular direction allow them to attain the large effective area required to match the observed brightnesses, despite their narrow widths.

We note that \citep{2010ApJ...708..232B} observed features out to a maximum delay of $\tau=1000$ to $1200\ \mu\mathrm{s}$;
if we suppose that these features at greatest delay have width of one Fresnel zone, 
the observations suggest a minimum radius for noodles of about 650\ km.
This is comparable to the ion inertial scale or the cyclotron radius in the scattering medium;
theory predicts that one or the other should be the minimum scale of a turbulent cascade.
Other observations have found similar values for a minimum size of the fluctuations responsible for scattering \citep{1990ApJ...353L..29S,2018ApJ...865..104J}.

\section*{Acknowledgements}

I thank Michael D. Johnson for essential discussions during the initiation of this work, and Walter Brisken for sharing his data.

%
%
%
\bibliographystyle{mnras}
\bibliography{wavebib} 



\bsp	
\label{lastpage}
\end{document}